\documentclass[aps,prb,floatfix,superscriptaddress,noshowpacs,twocolumn]{revtex4-1}
\usepackage{graphics,graphicx,epsfig,amsmath,amssymb,epstopdf,wasysym,comment}
\usepackage{hyperref}
\usepackage[dvipsnames]{xcolor}
\usepackage{braket,amsfonts}
\usepackage{dsfont}
\usepackage{hyperref}
\usepackage[dvipsnames]{xcolor}

\allowdisplaybreaks

\newcommand{\be}{\begin{equation}}
\newcommand{\ee}{\end{equation}}
\newcommand{\bea}{\begin{eqnarray}}
\newcommand{\eea}{\end{eqnarray}}
\newcommand{\ba}{\begin{eqnarray*}}
\newcommand{\ea}{\end{eqnarray*}}
\newcommand{\dagga}{{\phantom{\dagger}}}
\newcommand{\bR}{\mathbf{R}}
\newcommand{\bQ}{\mathbf{Q}}

\newcommand{\bq}{\mathbf{q}}

\newcommand{\bk}{\mathbf{k}}

\newcommand{\dis}{\displaystyle}

\newcommand{\up}{\uparrow}
\newcommand{\down}{\downarrow}
\newcommand{\fract}[2]{\frac{\dis \;#1\;}{\dis \;#2\;}}
\newcommand{\Tr}{\mathrm{Tr}}
\newcommand{\eqn}[1]{(\ref{#1})}

\newcommand{\ep}{{\epsilon}}

\newcommand{\bw}{\begin{widetext}}
\newcommand{\ew}{\end{widetext}}
\newenvironment{eqs}%
{\begin{equation} \begin{aligned}}%
{\end{aligned} \end{equation} }
\newcommand{\beal}{\begin{eqs}}
\newcommand{\eal}{\end{eqs}}
\newcommand{\bd}[1]{{\boldsymbol{#1}}}
\newcommand{\esp}[1]{\text{e}^{#1}}

\begin{document}
\title{Exciton topology and condensation in a model quantum spin Hall insulator
}

\author{Andrea Blason}
\affiliation{International School for
  Advanced Studies (SISSA), Via Bonomea
  265, I-34136 Trieste, Italy} 
\author{Michele Fabrizio}
\affiliation{International School for
  Advanced Studies (SISSA), Via Bonomea
  265, I-34136 Trieste, Italy} 

\begin{abstract}
We study by a consistent mean-field scheme the role on the single- and two-particle properties of a local electron-electron repulsion in the Bernevig, Hughes and Zhang model of a quantum spin Hall insulator. We find that the interaction fosters the 
intrusion between the topological and non-topological insulators of a new insulating and magnetoelectric phase that breaks spontaneously inversion and time reversal symmetries, but not their product. The approach to this phase from both topological and non-topological sides is signalled by the softening of two exciton branches, i.e., whose binding energy reaches the gap value, that possess, in most cases, finite and opposite Chern numbers, thus allowing this phase being regarded as a condensate of topological excitons. We also discuss how those excitons, and especially their surface counterparts, may influence the physical observables.

\end{abstract}
                         
\maketitle

\section{Introduction}
The physics of excitons in topological insulators has attracted considerable interest in the last decade, see, not as an exhaustive list, Refs.~\onlinecite{Moore-exciton-PRL2009,Pikulin-PRL2014,Budich-PRL2014,Fuhrman-exciton-PRL2015,Park6599,PhysRevLett.118.096604,Du-Nat-Comm-2017,Bi2Se3}, 
recently renewed \cite{Molinari-Nature-Nano-2020} by the evidence of a 
quantum spin Hall effect in two-dimensional transition metal dichalcogenides \cite{Qian-TMDC-Science2014,Tang-Nat-Phys-2017,Peng-Nat-Comm-2017}. \\
More precisely, a consistent part of the research activity has focused into the possibility of an exciton condensation in thin samples of topological insulators \cite{Moore-exciton-PRL2009,Efimkin-PRB2012,Franz-thin-PRB2012,Vignale-exciton-PRL2012,Pikulin-PRL2014,Budich-PRL2014,Du-Nat-Comm-2017}, 
much in the spirit of what was proposed \cite{Lozovik-BLG-JETP2008,
MacDonald-BLG-PRB2008} and observed \cite{Liu-BLG-Nat-Phys2017} in bilayer graphene. \\
In addition, the puzzling properties of the purported topological Kondo insulator SmB$_6$ \cite{PhysRevB.94.165154,Fisk-SmB6-PRB2019,Hartstein-SmB6-Nat-Phys2018}  
prompted interest in the excitons of such material \cite{Fuhrman-exciton-PRL2015,PhysRevB.92.085133,PhysRevB.94.235125,Park6599,PhysRevLett.118.096604,spinexciton}
as partly responsible for its anomalous behaviour. 

Even though evidences of excitons exist also in the three-dimensional topological insulator Bi$_2$Se$_3$ \cite{Bi2Se3}, 
besides those in the still controversial SmB$_6$, a systematic study in model topological insulators is largely lacking.~\cite{Garate-exciton-PRB2011,PhysRevB.92.085133,PhysRevB.96.161101,Galitski-exciton-PRB2018} Our aim here is to partly fill this gap. Specifically, we consider 
the prototypical model of a Quantum Spin Hall Insulator (QSHI) introduced by Bernevig, Hughes 
and Zhang \cite{Bernevig1757}, and add a local interaction compatible with the 
symmetries, which, e.g., allow for a dipole-dipole term. We deal with such an interaction in a conserving mean-field scheme. Namely, we assume the Hartree-Fock expression of the self-energy functional to compute the single-particle Green's function. Next, we calculate the excitons by solving the Bethe-Salpeter equations for the response functions, using as irreducible vertex the functional derivative of the Hartree-Fock self-energy functional with respect to the Green's function; what is often called random phase approximation plus exchange.~\cite{RPA+E} \\
Our main result is that, starting 
from the non-interacting QSHI, branches of excitons that transform into each other 
under time reversal detach from the continuum of particle-hole excitations, and gradually soften upon increasing interaction strength. When the latter exceeds a critical value, those excitons become massless, and thus condense through a second order critical point, which coincides with that obtained directly through the Hartree-Fock calculation not forcing any symmetry. Such symmetry broken phase is still insulating, and displays magnetoelectric effects. Upon further increasing 
interaction, it eventually gives in to the non-topological symmetry invariant 
insulator via another second order transition. None of those transitions is accompanied by any gap closing; therefore uncovering a path between the QSHI and the trivial insulator 
that does not cross any gapless point\cite{PhysRevB.83.245132,Ezawa-SciRep2013,PhysRevLett.120.186802}, thanks to the interaction-driven spontaneous breakdown of time reversal symmetry. \\
We also find that, approaching the excitonic insulator from the QSHI, the excitons themselves 
may acquire a non trivial topology signalled by a non-zero Chern number, suggestive of the existence of chiral exciton edge modes. In addition, we have evidences that, in open boundary geometries, exciton condensation occurs at the surface earlier than in the bulk, which also 
foresees the existence of non-chiral surface excitons that go soft before the bulk ones.
\cite{Shitade-PRL2009,PhysRevB.85.235449,Adriano-edge-PRB2017}\\
Our findings may have observable consequences that we discuss, some of which not in disagreement with existing experimental evidences.

This work is organised as follows. In section \ref{Model} we introduce the interacting model 
Hamiltonian, while the conserving Hartree-Fock approximation that we use to deal with interaction is discussed in sections \ref{HF-appx} and \ref{sec:BS}. 
The results of the calculations are presented in section \ref{Results}, specifically:   
in \ref{Results:A} the Hartree-Fock phase diagram; 
in \ref{Results:B} the excitons in the quantum spin Hall insulating phase; 
and, finally, in \ref{Results:C} the magnetoelectric effect in the excitonic insulator.  
Section \ref{Conclusions} is devoted to concluding remarks. 

\section{The model Hamiltonian}
\label{Model}
\begin{figure}[t]
\centering
\includegraphics[width=0.97\columnwidth]{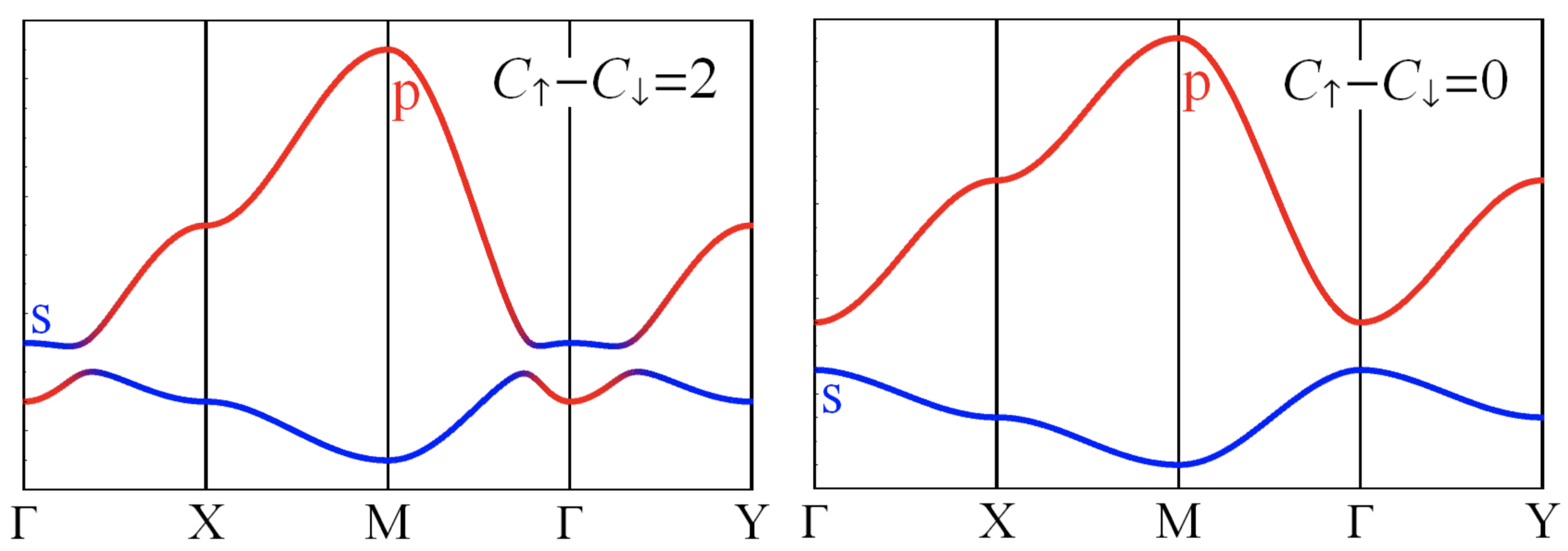}
\caption{Band structure of the BHZ model in the topological, left panel, and trivial, right panel, regimes. Blue and red colours indicate, respectively, even ($s$ orbital character) and odd ($p$ orbital character) parity under inversion.}
\label{Fig1}
\end{figure}

We shall consider the model introduced by Bernevig, Hughes and Zhang, 
after them named BHZ model, to describe the QSHI 
phase in HgTe quantum wells \cite{Bernevig1757}. The BHZ model involves two spinful Wannier orbitals per unit cell, which transform like $s$-orbitals, $\ket{s \,\sigma}$, where $\sigma=\up,\down$ refers to the projection of the spin along the $z$-axis, and like the $J=3/2$, $J_z=\pm 3/2$ spin-orbit 
coupled combinations of $p$-orbitals, i.e., 
\beal
\ket{p_x+ip_y \up} &=\ket{p_{+1} \up} \equiv \ket{p\up}\,, \\
\ket{p_x-ip_y \down} &=\ket{p_{-1}\down} \equiv \ket{p\down}\,.\label{p-orbitals}
\eal
We introduce two sets of Pauli matrices, $\sigma_a$ and $\tau_a$, $a=0,\dots,3$, with $a=0$ denoting the identity, which act, respectively, in the spin, $\up$ and $\down$, and orbital, $s$ and $p$, spaces.\\
With those definitions, the BHZ tight-binding Hamiltonian on a square lattice includes all  onsite potentials and nearest neighbour hopping terms  
that are compatible with inversion, time reversal and $\text{C}_4$ symmetry\cite{Rothe_2010}, and reads
\be
\mathcal{H}_0 = \sum_\bk  \bd{\Psi}^\dagger_\bk  \, \hat H_0(\bk) \,\bd{\Psi}^\dagga_\bk = \sum_{i j}  
\bd{\Psi}^\dagger_{i} \,  \hat H_0(\bR_i-\bR_j) \, \bd{\Psi}^\dagga_{j}
 , \label{BHZ-Ham}
\ee
at density corresponding to two electrons per site, where 
\beal
\bd{\Psi}^\dagga_\bk
&=\begin{pmatrix}
s^\dagga_{\bk\up}\\
s^\dagga_{\bk\down}\\
p^\dagga_{\bk\up}\\
p^\dagga_{\bk\down}
\end{pmatrix}\,,&
\bd{\Psi}^\dagga_i
&=\begin{pmatrix}
s^\dagga_{i\up}\\
s^\dagga_{i\down}\\
p^\dagga_{i\up}\\
p^\dagga_{i\down}
\end{pmatrix}\,,
\eal
are four component spinors in momentum, $\bd{\Psi}_\bk$, and real, 
$\bd{\Psi}_i$, space, with $i$ labelling the unit cell at position $\bR_i$. 
$\hat{H}_0(\bk)$ is the $4\times 4$ matrix
\beal
\hat H_0(\bk) &= \big(M- t\,  \ep_\bk \big) \,\sigma_0\otimes\tau_3 - t' \,\ep_\bk\,\sigma_0\otimes\tau_0   \\ 
 &\quad +\lambda\,\sin k_x\;\sigma_3\otimes\tau_1 
-\lambda\,\sin k_y\;\sigma_0\otimes\tau_2 \, ,\label{H_0}
\eal
with $\bk=\big(k_x,k_y\big)$ and $\ep_\bk=\big(\cos k_x +\cos k_y \big)$ while $\hat H_0(\bR_i-\bR_j)$ its Fourier transform in real space. 
The parameters $t'-t$, $t'+t$ and $\lambda$ correspond, respectively, 
to the $s-s$, $p-p$ and $s-p$ nearest neighbour hybridisation amplitudes. Finally, $M$ describes an onsite energy difference between the two orbitals.
\\
Hereafter, we shall analyse the Hamiltonian \eqn{H_0} for $M>0$, 
$t'=0.5\,t$ and $\lambda=0.3\,t$. The precise values of the latter two 
is not crucial to the physics of the model. What really matters is the relative magnitude 
of $M$ and $t$, and the finiteness of $\lambda$. Therefore, for sake of simplicity, 
we shall set $t=1$ as the unit of energy. \\
The band structure can be easily calculated and yields two bands, each 
degenerate with respect to the spin label $\sigma$;   
a conduction and a valence band, 
with dispersion $\ep_c(\bk)$ and $\ep_v(\bk)$, respectively, which read 
\beal
\ep_c(\bk) &= -t'\,\ep_\bk + E_\bk\,,& 
\ep_v(\bk) &= -t'\,\ep_\bk - E_\bk\,, \label{e-c-v}
\eal
where 
\beal
E_\bk &= \sqrt{\;\big(M-\ep_\bk\big)^2 + \lambda^2\,\sin^2 k_x +  
\lambda^2\,\sin^2 k_y\;}\;.\label{E_k}
\eal
With our choice of parameters, these bands describe a direct gap semiconductor
for any $M\not=2$. 
At the high symmetry points in the Brillouin Zone (BZ), the bands have a defined orbital character, i.e., a defined parity under inversion. Specifically,  at $\bd{\Gamma}=(0,0)$, 
\be
\ep_c(\bd{\Gamma}) = -2t' + \big|M-2\big|\,,\quad 
\ep_v(\bd{\Gamma}) = -2t' - \big|M-2\big|\,, \label{e-c-v-Gamma}
\ee
valence and conduction bands have, respectively, $s$ and $p$ orbital character
if $M<2$, and vice versa if $M>2$. On the contrary, at the zone boundary points $\bd{M}=(\pi,\pi)$, $\bd{X}=(\pi,0)$, and $\bd{Y}=(0,\pi)$,  
\beal
\!\!\ep_c(\bd{M}) &= 2t' + \big(M+2\big)\,,&\ep_v(\bd{M}) &= 2t' - \big(M+2\big)\,,  \\
\!\!\ep_c(\bd{X}) &=\ep_c(\bd{Y})=M \,, &\ep_v(\bd{X}) &= \ep_v(\bd{Y})=-M\,, 
\label{e-c-v-M}
\eal
the valence band is $p$ and the conduction one $s$ for any value of $M$. 
It follows that, if $M<2$, there is an avoided band crossing, due to $\lambda\not=0$, moving from $\bd{\Gamma}$ towards the BZ boundary, while, if $M>2$, each band has predominantly a single orbital character, $s$ the conduction band and $p$ the valence one, see Fig.~\ref{Fig1}. At $M=2$ the gap closes at $\bd{\Gamma}$, around which the dispersion becomes Dirac-like,  
\beal
\ep_{c}(\bk) &\simeq +\lambda\,\big|\bd{k}\big|\,,&
\ep_{v}(\bk) &\simeq -\lambda\,\big|\bd{k}\big|\,.
\eal
The transition between the two insulating phases is known to have topological 
character \cite{Bernevig1757}.
\\
We note that the Hamiltonian $\hat H_0(\bk)$ commutes with $\sigma_3$, i.e., is invariant 
under $U(1)$ spin rotations around the $z$-axis, as well as under inversion and time reversal, respectively represented by the operators 
\beal
\mathcal{I}:\,\hat{H}_0(\bk) &= \sigma_0\otimes\tau_3\;\hat{H}_0(-\bk)\;
\sigma_0\otimes\tau_3\,,\\
\mathcal{T}:\,\hat{H}_0(\bk) &=\big(-i\sigma_2\otimes\tau_0\big)\;\hat{H}^*_0(-\bk)\;
\big(i\sigma_2\otimes\tau_0\big)
 \,.
\eal
In addition, it is invariant under spatial, i.e., not affecting spins, $\text{C}_4$ rotations, which correspond to   
\beal
\mathcal{C}_4 &:\,\hat{H}_0(\bk) & = \esp{-i\,\frac{\pi}{2}\,L_3}\;
\hat{H}_0\big(\text{C}_4(\bk)\big)\;\esp{i\,\frac{\pi}{2}\,L_3}\,, \label{C_4}
\eal
where $\text{C}_4(\bk)=\big(k_y,-k_x\big)$, and the $z$-component of the 
angular momentum operator is  
\be
L_3 = \sigma_3\otimes\fract{\tau_0-\tau_3}{2}\,.\label{L_3}
\ee 
Evidently, since the Hamiltonian is also invariant under spin $U(1)$ rotations, with generator $S_3=\sigma_3\otimes\tau_0/2$, it is also invariant under $\pi/2$ rotations 
with generator the total angular momentum along $z$, i.e., $J_3=L_3+S_3$, which 
provides a better definition of $\text{C}_4$.   
 \\
We observe that, if $C_\sigma$ is the Chern number of the spin-$\sigma$ valence-band electrons, then 
invariance under both inversion and time reversal entails a vanishing 
$\big(C_\up+C_\down\big)$, which is proportional to the transverse 
charge-conductance, but a possibly non zero $\big(C_\up-C_\down\big)$, 
which would correspond to a finite transverse spin-conductance,  
thus the nontrivial topology of a QSHI.~\cite{PhysRevLett.95.226801} 
Specifically, $\big(C_\up-C_\down\big)\not=0$ occurs when $M<2$ \cite{Bernevig1757,PhysRevB.83.245132}, not surprisingly in view of the avoided band crossings. We emphasise that a robust topological invariant can be defined provided spin $U(1)$ symmetry is preserved. 

\noindent
So far we have discussed the main properties of the non-interacting Hamiltonian \eqn{BHZ-Ham}. However, physically, electrons unavoidably interact with each other. We shall therefore add to the non-interacting Hamiltonian \eqn{BHZ-Ham} a local Coulomb interaction $U_\text{int}$, thus neglecting its long range tail, which 
includes, besides monopoles terms, also a dipole-dipole interaction $U_\text{dip}$, which is here allowed by symmetry.  Specifically,
\beal
U_\text{int} = U + U_\text{dip} \, ,\label{U-int}
\eal 
where
\be
U =  \sum_{i} \Big(U_s \, n_{is\up}   n_{is\down}  +  U_p \, n_{ip\up}  n_{ip\down}  + V \, n_{is}  n_{ip}\Big) \, , \label{U}
\ee
includes monopole terms, while the dipole-dipole interaction, projected onto our basis of single-particle wavefunctions, reads
\beal
U_\text{dip} &= \fract{J}{2} \sum_i \, \bigg[\,\Big(\,\bd{\Psi}^\dagger_{i}\;
\sigma_0\otimes\tau_1\;\bd{\Psi}^\dagga_i\;\Big)^2\\
&\qquad\qquad\qquad
+ \Big(\,\bd{\Psi}^\dagger_{i}\;
\sigma_3\otimes\tau_2\;\bd{\Psi}^\dagga_i\;\Big)^2\;\bigg]
\,.\label{U-dip}
\eal
All coupling constants, $U_s$, $U_p$, $V$ and $J$, are positive, $n_{is\sigma}=s^\dagger_{i\sigma}\,s^\dagga_{i\sigma}$, 
$n_{ip\sigma}=p^\dagger_{i\sigma}\,p^\dagga_{i\sigma}$, and 
$n_{is(p)} = n_{is(p)\up}+n_{is(p)\down}$.
Hereafter, in order to reduce the number of independent parameters and thus 
simplify the analysis, we shall take $U_s=U_p=U$. Moreover, the numerical 
solution will be carried out with the further simplification $U=V$. 

We end mentioning that a calculation similar to the one we are going to present has been 
already performed by Chen and Shindou \cite{PhysRevB.96.161101}, though in the magnetised BHZ 
model, which includes just two orbitals, $\ket{s\,\up}$ and, differently from our time-reversal invariant case, see Eq.~\eqn{p-orbitals}, the $J=3/2$, $J_z=+1/2$ orbital $\ket{p_+\,\down}$.


\section{Hartree-Fock approximation}
\label{HF-appx}
The simplest way to include the effects of a not too strong interaction 
is through the Hartree-Fock approximation (HF), which amounts to 
approximate the self-energy functional simply by the Hartree and Fock 
diagrams. For sake of simplicity, we shall introduce the HF approximation 
under the assumption of unbroken translational symmetry, 
so that the lattice total momentum is a good quantum number. 
Whenever needed, we will mention what changes when translational symmetry is broken. 

Within the HF approximation, if 
\beal
\hat{G}_0(i\ep,\bk)^{-1} &= i\ep - \hat{H}_0(\bk)\,,\label{G_0}
\eal
is the inverse of the non-interacting $4\times 4$ Green's function matrix at momentum $\bk$ and in Matsubara frequencies, $i\ep$, the interacting Green's function 
is
\beal
\hat{G}(i\ep,\bk)^{-1} &= \hat{G}_0(i\ep,\bk)^{-1} -\hat{\Sigma}_{HF}\big[\hat{G}\big]\,,\label{Dyson}
\eal
where, in the specific case under consideration, the self-energy within the HF approximation is functional of the local Green's function
\be
\hat{\Sigma}_{HF}\Big[\hat{G}(\bR_i,\bR_i)\Big] = \sum_{\alpha,a=0}^3\,
\sigma_\alpha\otimes\tau_a\;\Gamma^0_{\alpha a}\,\Delta_{\alpha a}(\bR_i)\,,
\label{Dyson-1}
\ee
with
\beal
\Delta_{\alpha a}(\bR_i) &\equiv T\sum_\ep\,\text{e}^{i\ep\,0^+}\;
\text{Tr}\Big(\hat{G}(i\ep,\bR_i,\bR_i)\,\sigma_\alpha\otimes\tau_a\Big)\\
&= \braket{\,
\bd{\Psi}^\dagger_i\;\sigma_\alpha\otimes\tau_a\;
\bd{\Psi}^\dagga_i\;} \equiv \braket{\,\text{O}_{\alpha a}(\bR_i)\,}
\in \mathbb{R}\;,\label{Dyson-2}
\eal
which become independent of the site coordinates $\bR_i$ if translational symmetry holds, i.e., $\Delta_{\alpha a}(\bR_i)\to \Delta_{\alpha a}$, 
$\forall\,\bR_i$. 
The Dyson equation \eqn{Dyson}, together with \eqn{Dyson-1} and
\eqn{Dyson-2}, yield a self-consistency condition that has to be solved. 
$\Gamma^0_{\alpha a}$ are the irreducible scattering amplitudes  
in the HF approximation, and, through Eq.~\eqn{U-int}, their expressions can be readily derived:
\beal
\Gamma^0_{00}&= \fract{U+2V-2J}{4}\,,& \Gamma^0_{03}&= \fract{U-2V+2J}{4}\,,\\
\Gamma^0_{01}&= -\fract{V-4J}{4}\,,&
\Gamma^0_{02}&= -\fract{V}{4}\,,\\
\Gamma^0_{10}&= \Gamma^0_{20}= -\fract{U}{4}\,,&
\Gamma^0_{30}&= -\fract{U+2J}{4}\,,\\
\Gamma^0_{11}&= \Gamma^0_{21}= -\fract{V+2J}{4}\,,&
\Gamma^0_{31}&= -\fract{V}{4}\,,\\
\Gamma^0_{12}&= \Gamma^0_{22}= -\fract{V-2J}{4}\,,&
\Gamma^0_{32}&= -\fract{V-4J}{4}\,,\\
\Gamma^0_{13}&= \Gamma^0_{23}= -\fract{U}{4}\,,&
\Gamma^0_{33}&= -\fract{U-2J}{4}\,.\label{scattering-amplitudes}
\eal
We note that the scattering amplitudes posses the same spin $U(1)$ symmetry 
of the non-interacting Hamiltonian, namely, $\Gamma^0_{1a}=\Gamma^0_{2a}\not=
\Gamma^0_{3a}$, $a=0,\dots,3$. \\

The expectation values $\Delta_{00}(\bR_i)
=\braket{ n_{is}+n_{ip}}$ and $\Delta_{03}(\bR_i)
=\braket{ n_{is}-n_{ip}}$, which measure the local density 
and orbital polarisation, respectively, are  
finite already in absence of interaction. In this case, the effects of the scattering amplitudes $\Gamma^0_{00}$ and $\Gamma^0_{03}$ treated within HF are, respectively, to shift the chemical potential, 
which we can discard since we work at fixed density, and renormalise upward 
the value of $M$, thus enlarging the stability region of the non topological phase. \\
On the contrary, all other expectation values 
$\Delta_{\alpha a}(\bR_i)$, $(\alpha,a)\not=(0,0),(0,3)$, break one or more symmetries of the non-interacting Hamiltonian, and therefore vanish identically in the non interacting case. They could become finite should interaction be strong enough to lead to spontaneous symmetry breaking. 
We expect this should primarily occur in those channels 
whose scattering amplitudes are the most negative ones, being $\Delta_{\alpha a}(\bR_i)$ real by definition. If $V\simeq U$, as we shall assume in the following numerical calculations, the dominant symmetry breaking channels 
are therefore those with $(\alpha,a) = (3,0),\,(1,1),\,(2,1)$. We emphasise that the dipolar coupling constant $J$ plays an important role in splitting the large degeneracies 
of the scattering amplitudes in \eqn{scattering-amplitudes} that exist at $J=0$.\\
Specifically, 
\be
\Delta_{30}(\bR_i) = \braket{\bd{\Psi}_i^\dagger\,
\sigma_3\otimes\tau_0\,\bd{\Psi}_i^\dagga} 
=\sum_{l=s,p}\,\braket{\,n_{il\up}-n_{il\down}\,}\,, 
\label{Delta-AFM}
\ee 
corresponds to a spontaneous spin polarisation along the $z$-axis, which breaks time reversal symmetry. 
We shall investigate two possible magnetic orders, $\Delta_{30}(\bR_i)=\Delta_{30}\,\text{e}^{i\bd{Q}\cdot\bR_i}$, with $\bQ=(0,0)$ or $\bQ=(\pi,\pi)$, corresponding, respectively, to ferromagnetism or antiferromagnetism. We point out that the latter implies a breakdown of translational symmetry, in which case the Green's function is not anymore diagonal in $\bk$, but depends on it as well as on $\bk+\bQ$, so it becomes 
an $8\times 8$ matrix, and Eq.~\eqn{Dyson} must be  
modified accordingly.\\
On the contrary,  
\beal
\Delta_{1 1}(\bR_i) &= \braket{\bd{\Psi}_i^\dagger\,
\sigma_1\otimes\tau_1\,\bd{\Psi}_i^\dagga} \\
&= \sum_{\sigma=\up,\down}
\braket{ s^\dagger_{i\sigma}\,p^\dagga_{i-\sigma} + p^\dagger_{i\sigma}\,s^\dagga_{i-\sigma}}\,,\\
\Delta_{2 1}(\bR_i) &= \braket{\bd{\Psi}_i^\dagger\,
\sigma_2\otimes\tau_1\,\bd{\Psi}_i^\dagga} \\
&= -i\,\sum_{\sigma=\up,\down}\,\sigma\,
\braket{ s^\dagger_{i\sigma}\,p^\dagga_{i-\sigma} + p^\dagger_{i\sigma}\,s^\dagga_{i-\sigma}}\,,\label{Delta_11_12}
\eal
describe a spin-triplet exciton condensate polarised in the plane. 
Since the insulator has a direct gap, excitons condense at 
$\bQ=\bd{0}$, namely $\Delta_{\alpha 1}(\bR)=\Delta_{\alpha 1}$, $\forall\,\bR$ and $\alpha=1,2$. Moreover, because $\Gamma^0_{11}=\Gamma^0_{21}$, if we write
\beal
\Delta_{11} &= \Delta\,\cos\phi\,,&  \Delta_{21} &= \Delta\,\sin\phi\,,
\label{Delta}
\eal
we expect to find a solution with the same amplitude $\Delta$ for any value of $\phi$, which reflects the spin $U(1)$ symmetry. At any given $\phi$, such exciton condensation would break spin $U(1)$, inversion and time reversal symmetry. \\  
The emergence of an exciton condensate is therefore accompanied by a spontaneous spin $U(1)$ symmetry breaking. As previously mentioned, such breakdown prevents the existence of the strong topological invariant that  characterises the QSHI phase. Specifically, since the $z$-component of the spin is not anymore a good quantum number, the counter propagating edge states of opposite spin are allowed to couple each other, which turns their crossing into an avoided one.~\cite{PhysRevB.83.245132} The boundary thus becomes insulating, spoiling the topological transport properties of the QSHI. \\
We shall study this phenomenon performing an HF calculation in a ribbon geometry with open boundary conditions (OBC) along $x$, but periodic ones along $y$. Consequently, the non-interacting BHZ Hamiltonian looses 
translational invariance along the $x$-direction, while keeping it along $y$, so that the Green's function becomes a $4N_x\times 4N_x$ matrix for each momentum $k_y$, with 
$N_x$ the number of sites along $x$. A further complication is that HF self-energy in Eq.~\eqn{Dyson}  unavoidably depends on the $x$-coordinate of each site, which enlarges the number of self-consistency equations to be fulfilled. 
However, since those equations can be easily solved iteratively, we can still numerically afford ribbon widths, i.e., values of $N_x$, 
which provide physically sensible results with negligible size effects.  
\\
The OBC calculation gives access not only to the states that may form at the boundaries, but also, in the event of a spontaneous symmetry breaking, to the behaviour of the corresponding order parameter moving from the edges towards the bulk interior. In practice, 
we shall investigate such circumstance only in the region of Hamiltonian parameters when the dominant instability is towards the spin-triplet exciton condensation.

\section{Bethe-Salpeter equation}
\label{sec:BS}

If we start from the QSHI, $M<2$, and adiabatically switch on the interaction \eqn{U-int}, we expect that such phase will for a while survive because of the gap, till, for strong enough interaction, it will eventually give up to a different phase. We already mentioned that the first effect of interaction is to renormalise upward $M$, thus pushing the topological insulator towards the transition into the non topological one. Beside that, 
a repulsive interaction can also bind across gap particle-hole excitations, 
i.e., create excitons.\\
A direct way to reveal excitons is through the in-gap 
poles of linear response functions. Within the HF approximation for the 
self-energy functional, the linear response functions must be calculated solving the corresponding Bethe-Salpeter (BS) equations using the HF Green's functions together with the irreducible scattering amplitudes in Eq.~\eqn{scattering-amplitudes}, which are actually the functional derivatives of $\hat\Sigma_{HF}[\hat G]$  with respect to $\hat G$. \\
If the interaction is indeed able to stabilise in-gap excitons, 
their binding energy must increase with increasing interaction strength. 
It is therefore well possible that the excitons touch zero energy at a critical interaction strength, which would signal an instability towards a different, possibly symmetry-variant phase. Consistency of our approximation requires that such instability should also appear in the unconstrained HF calculation as a transition from the topological insulator to another phase, especially if such transition were continuous. We shall check that is indeed the case. \\

\noindent
With our notations, see Eqs.~\eqn{Dyson-2} and \eqn{scattering-amplitudes}, 
a generic correlation function will be defined as
\beal
\chi_{\alpha a;\beta b}(\tau,\bR) &\equiv 
- \big\langle\,T_\tau\Big(\,\text{O}_{\alpha a}(\tau,\bR)\,
\text{O}_{\beta b}(0,\bd{0})\,\Big)\,\big\rangle\,,
\eal  
where $T_\tau$ is the time-ordering operator, and the operators 
$\text{O}_{\alpha a}(\bR_i)=\bd{\Psi}_i^\dagger\sigma_\alpha\otimes\tau_a
\bd{\Psi}_i$ are evolved 
in imaginary time $\tau$. Spin $U(1)$ symmetry implies that the $z$-component 
$S_z$ of the total spin is conserved. It follows that the only non zero 
correlation functions $\chi_{\alpha a;\beta b}$ have 
$\alpha$ and $\beta$ either 0 and 3, corresponding to $S_z=0$, 
or 1 and 2, satisfying   
\beal
\chi_{1 a;1 b}(\tau,\bR) &=\chi_{2 a;2 b}(\tau,\bR)\,,\\
\chi_{1 a;2 b}(\tau,\bR) &=-\chi_{2 a;1 b}(\tau,\bR)\,,\label{symm-chi}
\eal
whose combinations $\chi_{1 a;1 b} \pm i\,\chi_{1 a;2 b}$ describe the independent propagation of 
$S_z=\pm1$ particle-hole excitations. \\
The Fourier transform $\chi_{\alpha a;\beta b}(i\omega,\bq)$,  
in momentum and in Matsubara frequencies, are obtained through the solution of a set of BS equations 
\beal
&\chi_{\alpha a;\beta b}(i\omega,\bq) = \chi^{(0)}_{\alpha a;\beta b}(i\omega,\bq) \\
&\qquad\qquad\quad + \sum_{\gamma c}\, \chi^{(0)}_{\alpha a;\gamma c}(i\omega,\bq)\;\Gamma^0_{\gamma c}\;\chi_{\gamma c;\beta b}(i\omega,\bq)\,,
\label{BS-chi}
\eal
where 
\beal
&\chi^{(0)}_{\alpha a;\beta b}(i\omega,\bq) = 
\fract{1}{N}\sum_\bk\,T\sum_\ep\\
&\quad
\Tr\bigg(\sigma_\alpha\otimes\tau_a\,\hat{G}(i\ep+i\omega,\bk+\bq)\,
\sigma_\beta\otimes\tau_b\,\hat{G}(i\ep,\bk)\bigg)\,.\label{chi_0}
\eal
In Eq.~\eqn{chi_0},
$N$ is the number of sites, and $\hat{G}(i\ep,\bk)$ the HF Green's function matrices.  \\

We shall perform the above 
calculation at zero temperature without allowing in the HF 
calculation any symmetry breaking. With this assumption, the HF Green's function reads   
\be
\hat{G}(i\ep,\bk) = \fract{(i\ep+t'\ep_\bk)\;\sigma_0\otimes\tau_0
+\hat H_{HF}(\bk)}{\;\big(i\ep-\ep_c(\bk)\big)
\big(i\ep-\ep_v(\bk)\big)\;}\\
\ee
where $\hat H_{HF}(\bk)$, $\ep_c(\bk)$ and $\ep_v(\bk)$ are 
those in equations~\eqn{H_0} and \eqn{e-c-v}, with $M$ in \eqn{H_0} and \eqn{E_k} replaced by an   
effective $M_{HF}$ determined through the self-consistency equation 
\beal
M_{HF} &= M - \fract{2\,\Gamma^0_{03}}{N}\sum_\bk\,\fract{\;M_{HF}-\ep_\bk\;}
{E_\bk}\;.\label{M_eff}
\eal
For $V\simeq U$, $\Gamma^0_{03}<0$ so that, since the sum over $\bk$ is positive, $M_{HF}>M$, as anticipated. \\

In short notations, and after analytic continuation on the real axis from above, $i\omega\to \omega + i\eta$, with $\eta>0$ infinitesimal, the physical 
response functions are obtained through the set of linear equations
\beal
\hat{\chi}(\omega,\bq) &= 
\bigg[\; \mathds{1} - \hat{\chi}^{(0)}(\omega,\bq)\;\hat{\Gamma}{\,^0}\;\bigg]^{-1}\,
\hat{\chi}^{(0)}(\omega,\bq)\,.
\eal
The excitons are in-gap solutions $\omega_i(\bq)$, i.e., 
\beal
\omega_i(\bq) < \omega_\text{min}(\bq) \equiv 
\min_{\bk}\Big(\ep_c(\bk+\bq)-\ep_v(\bk)\Big)\,,\label{omega_i omega_min}
\eal
of the equation
\beal
\text{det}\bigg[\; \mathds{1} - \hat{\chi}^{(0)}\big(\omega_i(\bq),\bq\big)\;\hat{\Gamma}{\,^0}\;\bigg]=0\,,
\eal
and have either $z$-component of the spin $S_z=0$, if they appear in the channels with 
$\alpha,\beta=0,3$, or $S_z=\pm 1$, in the channels with $\alpha,\beta=1,2$. 
For $\omega\simeq \omega_i(\bq)$, the response functions can be expanded in Laurent 
series \cite{PhysRevB.96.161101}
\be
\hat\chi(\bq,\omega) = \sum_i\,\frac{A_i(\bq)}{\omega-\omega_i(\bq)+i\eta} \ket{\psi_{i}(\bq)} \bra{\psi_i(\bq)} \,+ ...\,, 
\ee
where $\ket{\psi_{i}(\bq)}$ is the exciton wavefunction and $A_i(\bq)$ its spectral weight. This allows computing the exciton Chern number through the integral of the Berry curvature
\beal \label{eq-Chern}
C_i &= \frac{1}{2\pi} \int d^2\bq \; \, \Omega_i(\bq) \, , \\
\Omega_i(\bq) &= \text{Im} \, \braket{\, \partial_x \psi_i(\bq) \mid \partial_y \psi_i(\bq) \,} \, .
\eal
The curvature is even under inversion and odd under time reversal; if a system is invariant under both, the Chern number thus vanishes by symmetry.\\ 
We observe that all the excitons are invariant under inversion, but, 
while the $S_z=0$ ones are also invariant under time reversal, the latter maps onto each other the $S_z=+1$ and $S_z=-1$ excitons. Accordingly, only the $S_z=\pm 1$ excitons can have non-zero Chern numbers, actually opposite for opposite $S_z$, while the $S_z=0$ excitons are constrained to have trivial topology. 
We stress that such result, being based only upon symmetry considerations, remains valid for every inversion symmetric QSHI, and not only in the context of the interacting BHZ model.
\\
The exciton Chern number does not seem to be directly related to any  quantised observable. Nonetheless, as pointed out in Refs.~\onlinecite{PhysRevB.83.125109, PhysRevB.96.161101}, a nonzero $C_i$ ensures the presence of chiral exciton modes localised at the edges of the sample, which may have direct experimental consequences.

\section{Results}
\label{Results}

In the preceding sections we have introduced a conserving mean-field scheme to consistently calculate both the phase diagram and the linear response functions. Now, we 
move to present the numerical results obtained by that method at 
zero temperature and with Hamiltonian parameters $t'=0.5$, $\lambda=0.3$, $V=U=U_s=U_p$ 
and $J=U/16$, see equations~\eqn{H_0}, 
\eqn{U-int}, \eqn{U} and \eqn{U-dip}. The value of $J$ is estimated from 1s and 2p hydrogenoic orbitals, which may provide a reasonable estimate of the relative order of magnitude between the dipole interaction and the monopole one.

\subsection{Hartree-Fock phase diagram}
\label{Results:A}

\begin{figure}[htb]
\centerline{\includegraphics[width=0.7\columnwidth]{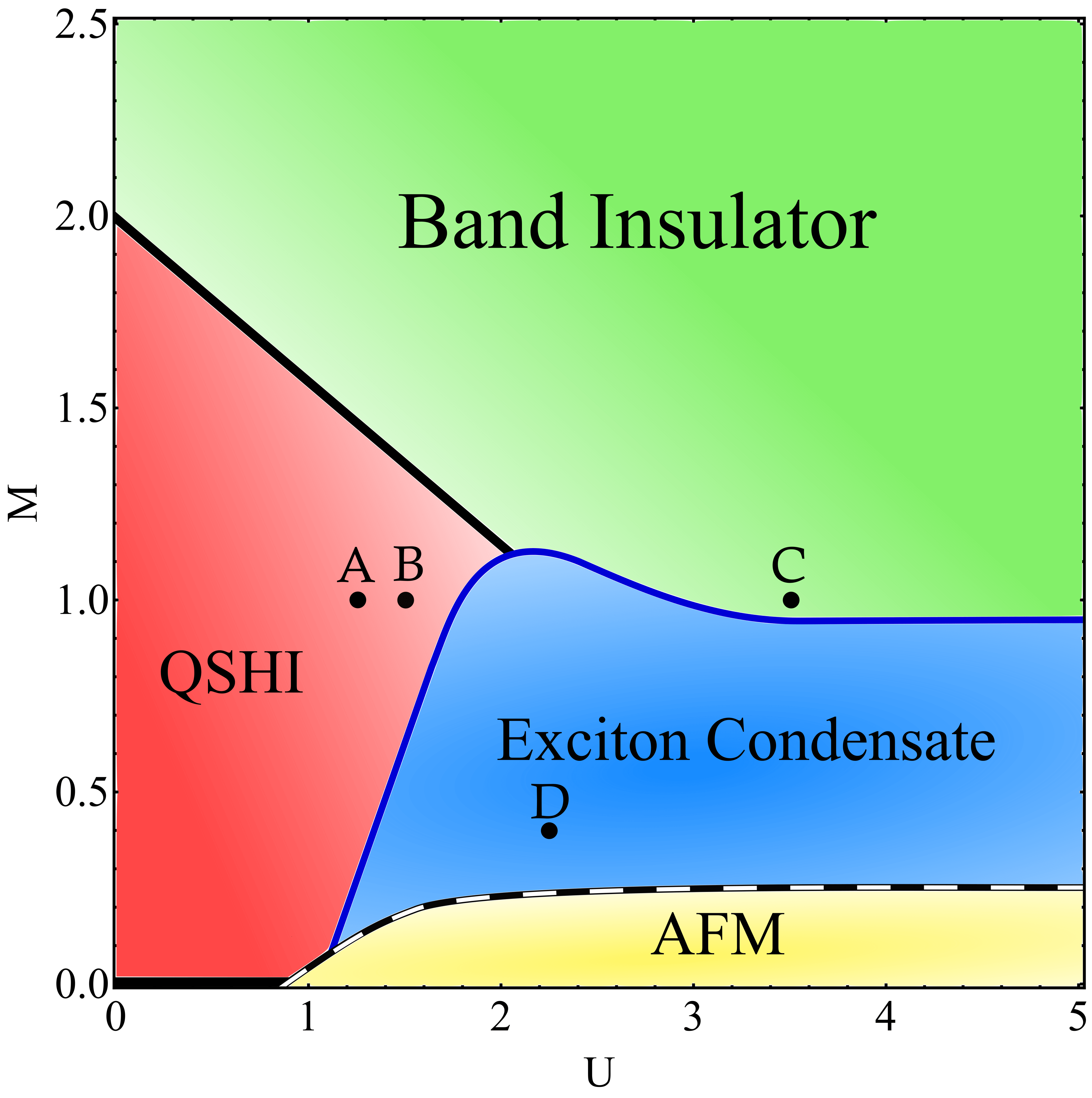}}
\caption{Hartree-Fock phase diagram at $\lambda=0.3$, $t'=0.5$, $U_s=U_p=V=U$ and $J=U/16$. The topological insulator is denoted as QSHI, while the non topological one as Band Insulator. For small value of $M$, antiferromagnetism (AFM) is stabilised upon increasing $U$. For larger values of $M$, $U$ stabilises a symmetry broken phase with Exciton Condensate. The thick black line that separates the QSHI from the Band Insulator, as well as that at $M=0$ extending from $U=0$ to the AFM phase, indicate a gapless metallic phase. The transition between the 
Exciton Insulator and the QSHI or the Band Insulator is continuous, while the transition into the AFM insulator is first order.}
\label{Fig2}
\end{figure}

The HF phase diagram is shown in Fig.~\ref{Fig2}.
\begin{figure}[thb]
\centerline{\includegraphics[width=0.8\columnwidth]{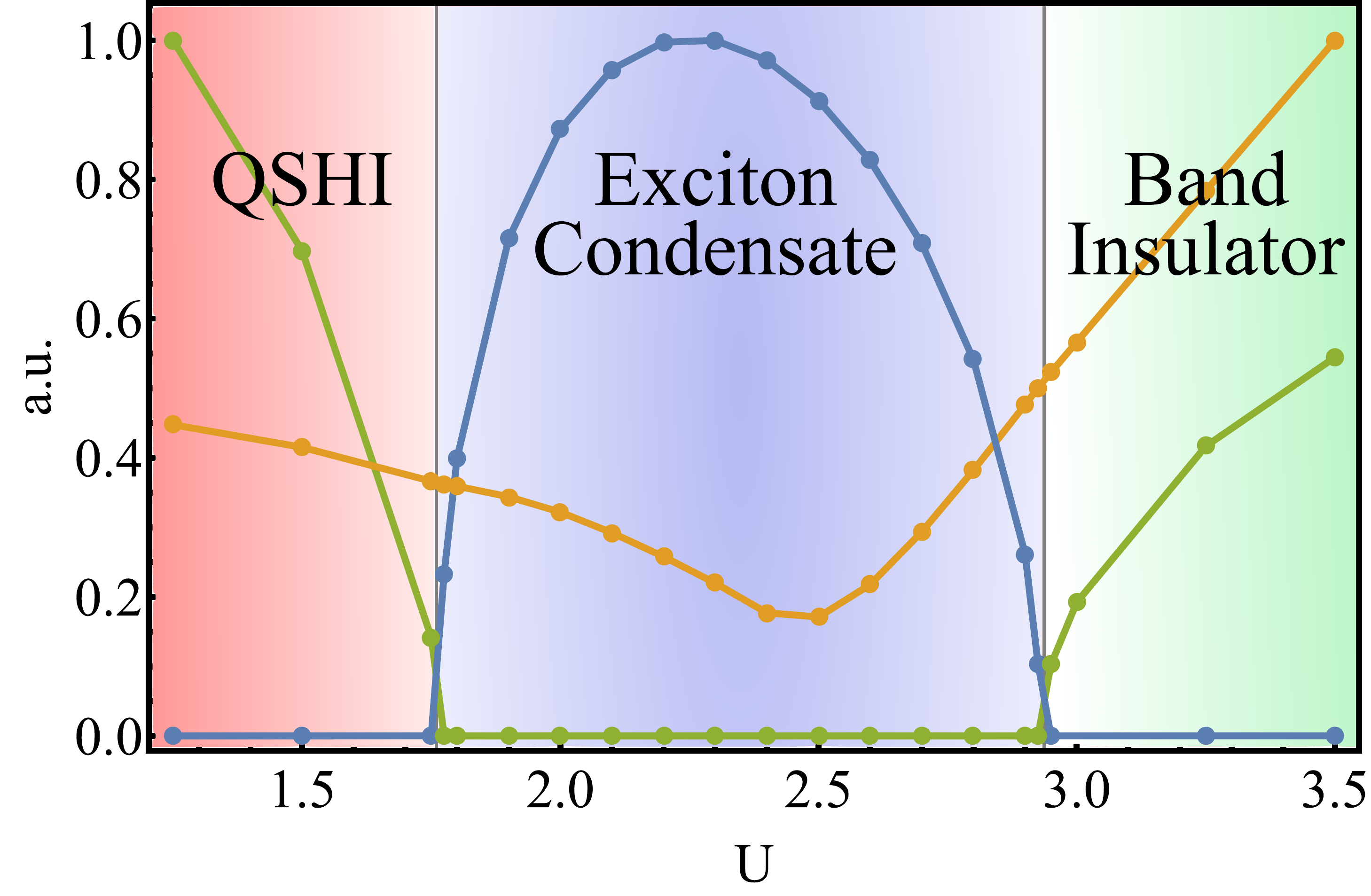}}
\vspace{-0.2cm}
\caption{Order parameter $\Delta$ of Eq.~\eqn{Delta} (blue), lowest $S_z=\pm1$ exciton energy at $\bQ=\bd{0}$ (green), and band gap (orange) along the path 
A to C in Fig.~\ref{Fig2} ($M=1$, $U\in[1.25,\,3.5]$), i.e.,  
from the topological to the trivial insulator crossing the exciton insulator. We note that the intermediate phase emerges exactly when the exciton becomes massless, as well as that the band gap never vanishes.}
\label{Fig3}
\end{figure}
As we previously mentioned, the interaction effectively increases $M$, thus pushing the transition from the topological 
insulator (QSHI) to the non topological one (Band Insulator) to lower values of $M$ the larger $U$. This is precisely what happens for $M\gtrsim 1.1$: 
$U$ increases the effective $M_{HF}$, see eq.~\eqn{M_eff}, until 
$M_{HF}=2$. At this point the gap closes and, for still larger $U$, the QSHI turns into the trivial Band Insulator. \\
For very small $M\lesssim 0.2$, upon increasing $U$ the QSHI gives in to an antiferromagnetic insulator (AFM), characterised by finite order parameters  
$\Delta_{30}(\bR_i)=\Delta_{30}\,\text{e}^{i\bd{Q}\cdot\bR_i}$, see Eq.~\eqn{Delta-AFM}, with $\bQ=(\pi,\pi)$, thus magnetised along $z$. HF predicts such transition to be of first order, in accordance to more accurate dynamical mean field theory calculations \cite{PhysRevB.98.045133}, which also explains why we do not find any precursory  
softening of $S_z=0$ exciton at $\bQ$.

More interesting is what happens for $0.2\lesssim M\lesssim 1.1$. Here, increasing the interaction $U$ drives a transition into a phase characterised by the finite order parameter
in Eq.~\eqn{Delta}, thus by a spontaneous symmetry breaking of  spin $U(1)$, time reversal and inversion symmetry. The breaking of time reversal allows the system moving from the QSHI to the Band Insulator without any gap closing \cite{PhysRevB.83.245132,Ezawa-SciRep2013,PhysRevLett.120.186802}, see Fig.~\ref{Fig3}. 
We note that the transition into the symmetry variant phase happens to be continuous, at least within HF. As we mentioned, consistency of our approach implies 
that this transition must be accompanied by the softening of the excitons whose condensation signals the birth of the symmetry breaking. These excitons are those with $S_z=\pm 1$, and indeed get massless on both sides of the transition, see Fig.~\ref{Fig3}. \\
\noindent

\begin{figure}[thb]
\vspace{-0.7cm}
\centering
\includegraphics[width=0.95\columnwidth]{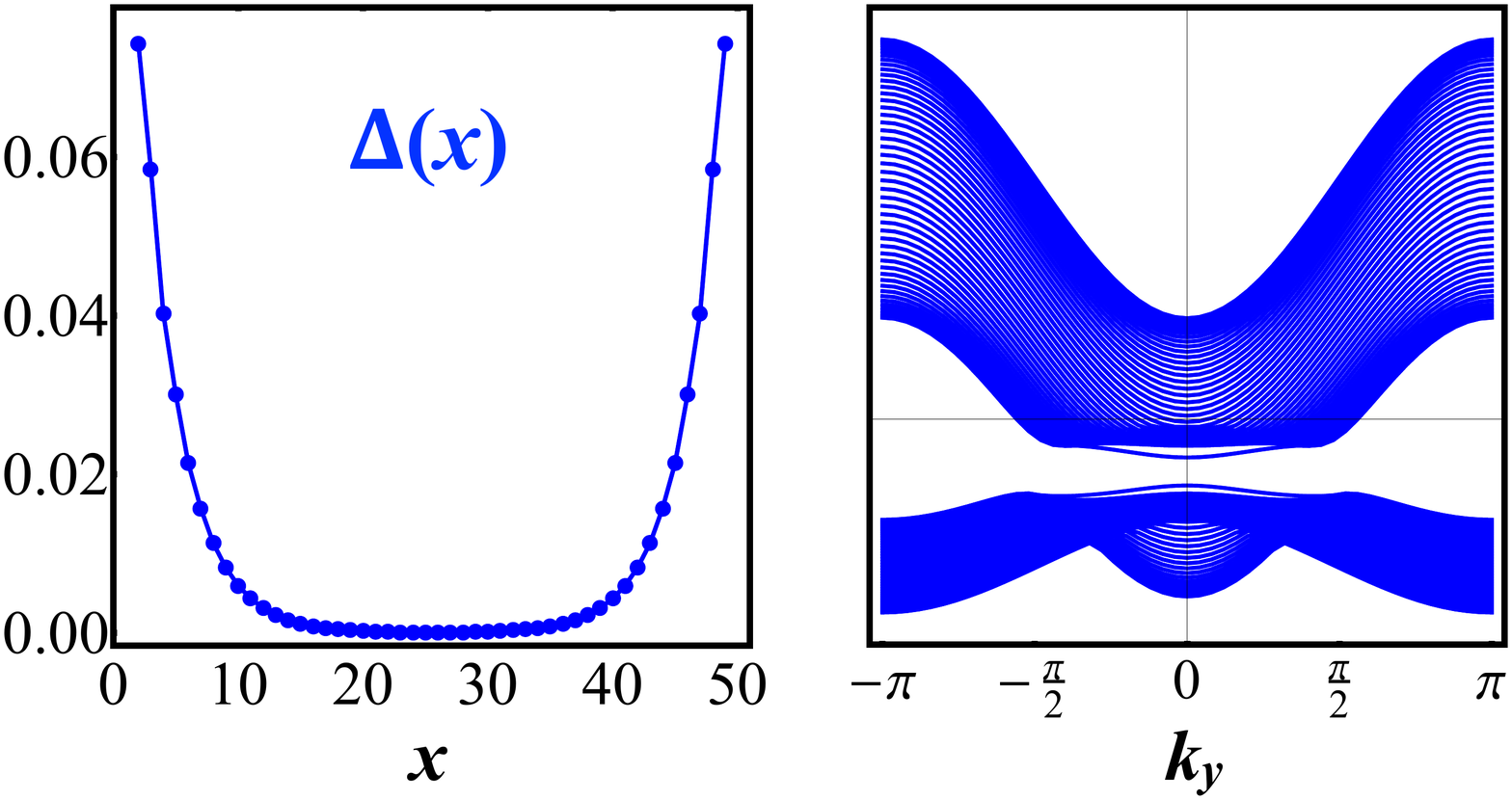}
\vspace{-0.7cm}
\caption{Left panel: exciton condensation order parameter $\Delta(x)$ in Eq.~\eqn{Delta} 
as function of the $x$-coordinate in a ribbon geometry with $N_x=50$ sites, calculated at  
point B in Fig.~\ref{Fig2} ($M=1$, $U=1.5$). Right panel: The ribbon band structure as function of the momentum $k_y$. We note that, even though the condensate is exponentially localised at the edges of the system, still it has a strong effect on the single-particle  edge states: a gap opens between the two branches, preventing topological spin transport.}
\label{Fig4}
\end{figure}

The HF numerical results in the ribbon geometry with OBC along $x$ show that electron correlations get effectively enhanced near the boundaries,
\cite{Shitade-PRL2009,PhysRevB.85.235449,Adriano-edge-PRB2017} unsurprisingly because of the reduced coordination.\cite{PhysRevLett.102.066806} Indeed, the order parameter is rather large at the edges, and, moving away from them, decays exponentially towards its bulk value, as expected in an insulator. Remarkably, even when the bulk is in the QSHI stability region, a finite symmetry breaking order parameter exponentially localised at the surface layer may still develop, see Fig.~\ref{Fig4} that refers to the point B in the phase diagram of Fig.~\eqn{Fig2}. In the specific two dimensional 
BHZ model that we study, such phenomenon is an artefact of 
the Hartree-Fock approximation, since the spin $U(1)$ symmetry cannot be 
broken along the one dimensional edges. Nonetheless, the enhanced quantum fluctuations, while preventing a genuine symmetry breaking, should all the same substantially affect the physics at the edges.      

We end the discussion of the Hartree-Fock phase diagram by comparing our results with those 
obtained by Xue and MacDonald \cite{PhysRevLett.120.186802}. These authors, too, apply  
the HF approximation to study the BHZ model but in the continuum limit and in presence of a long range Coulomb interaction.  They also find 
a path between the topological insulator and the trivial one that crosses 
another insulating phase characterised by spontaneous time reversal symmetry breaking, which, 
they argue, further breaks $\text{C}_4$ symmetry, thus being nematic. The HF band structure that we find in 
the exciton condensate phase is instead perfectly $\text{C}_4$ invariant, which might apparently indicate that our phase and that of Ref.~\onlinecite{PhysRevLett.120.186802} are different. In reality, we believe the two phases 
are just the same phase. Indeed, while it is true that the order parameter \eqn{Delta} 
is not invariant under the $\text{C}_4$ symmetry of Eq.~\eqn{C_4} that changes 
$\phi\to \phi-\pi/2$, such shift can be reabsorbed by a $-90^\circ$ spin $U(1)$ rotation. 
In other words, the order parameter \eqn{Delta} is invariant under a magnetic 
$\text{C}'_4$ symmetry of the Hamiltonian, whose generator of $\pi/2$ rotations is $L_3-S_3$ 
times the rotation of $\bk$. Due to such residual symmetry, the band structure does not show nematicity, as well as the magnetoelectric tensor we shall discuss later in section \ref{Results:C}.

\subsection{Excitons and their topological properties}
\label{Results:B}

The mechanism that triggers exciton topology is similar to the band inversion in the single-particle case: a topological exciton is composed by particle-hole excitations that have different parity under inversion in different regions of the BZ. In our case study, four possible orbital channels 
$\tau_a$, $a=0,\dots,3$, are allowed, each possessing a well defined parity: $\tau_1$ and $\tau_2$ odd, while $\tau_0$ and $\tau_3$ even. 
In the non-topological insulator, the $S_z=\pm 1$ excitons have the same parity character at 
all inversion invariant $\bk$-points, $\bd{\Gamma}$, $\bd{M}$, $\bd{X}$ and $\bd{Y}$, 
and thus are topologically trivial. 
On the contrary, in the QSHI, the highly mixed bands 
entail that all channels have finite weight in the exciton, which  
may acquire non trivial topology when its symmetry under parity changes among the inversion  invariant $\bk$-points, thus entailing 
one or more avoided crossings.
\\

\begin{figure}[t]
\includegraphics[width=0.65\columnwidth]{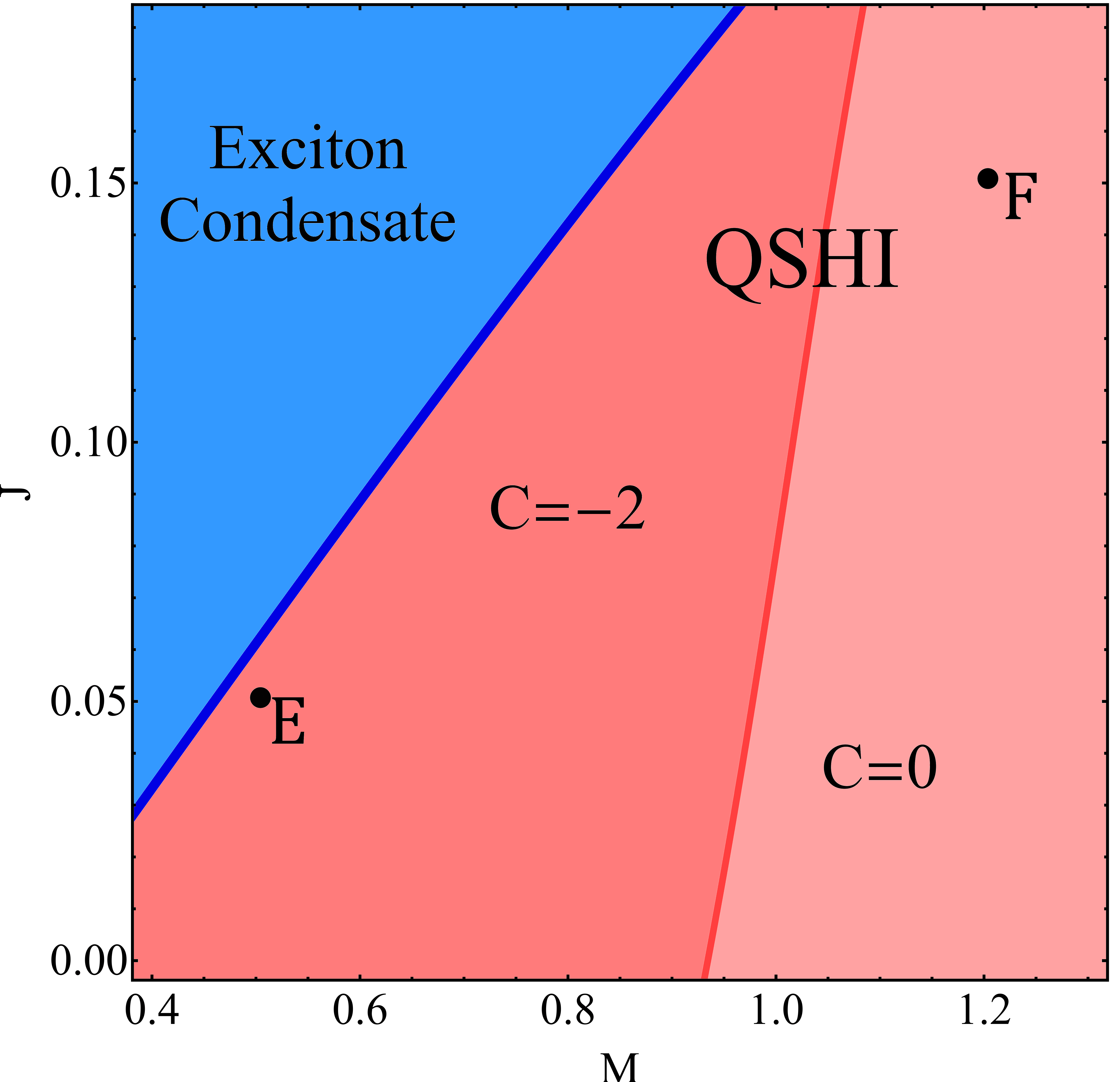}
\caption{Chern number of the most bound exciton with $S_z=-1$ at $\lambda=0.3$, $t'=0.5$, $U_s=U_p=V=1.5$, as function of $M$ and $J$ close to the transition from the QSHI 
to the symmetry broken phase.}
\label{Fig5}
\end{figure}

\begin{figure}[hbt]
\centering
\includegraphics[width=0.7\columnwidth]{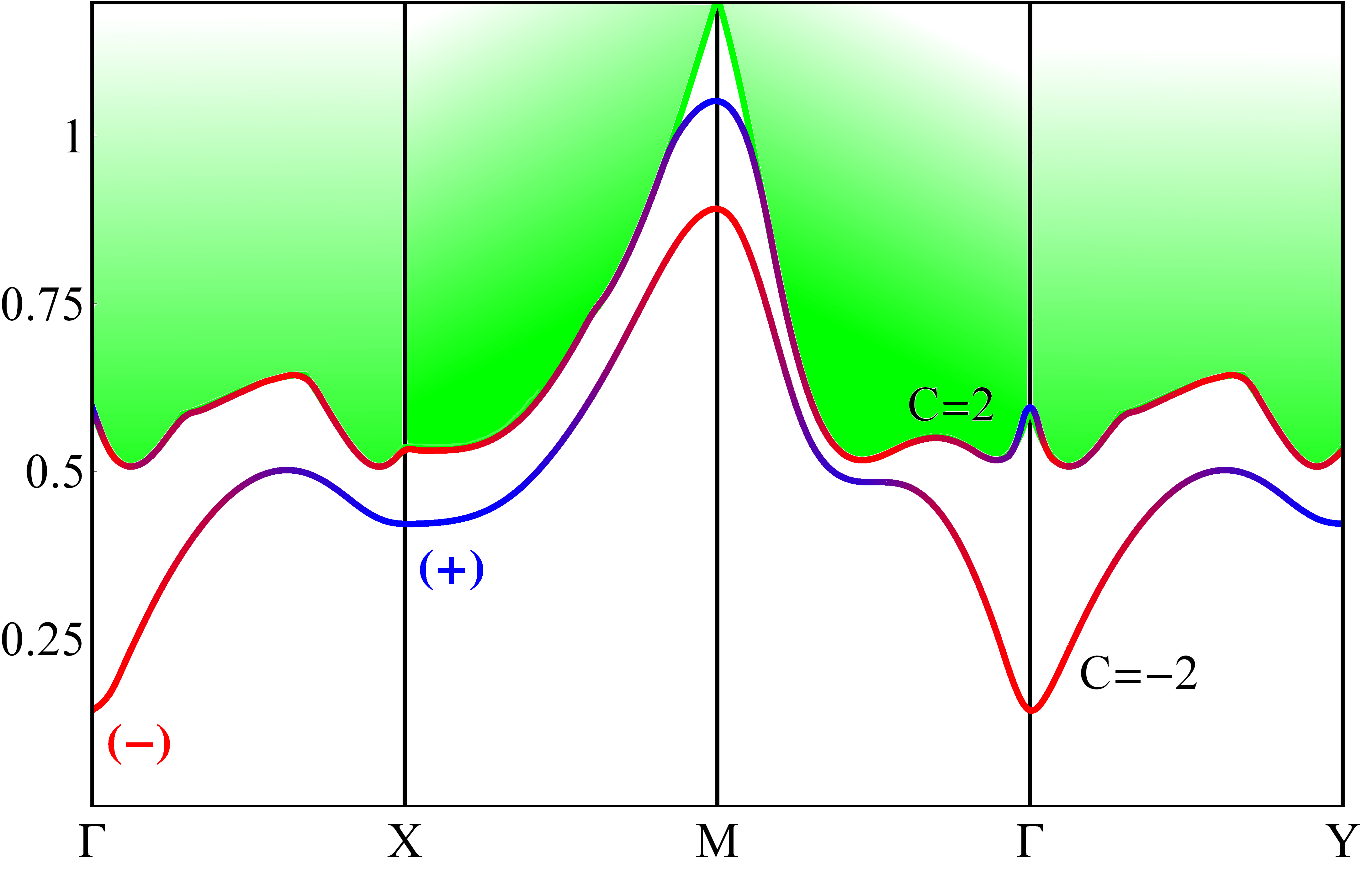}\\
\includegraphics[width=0.7\columnwidth]{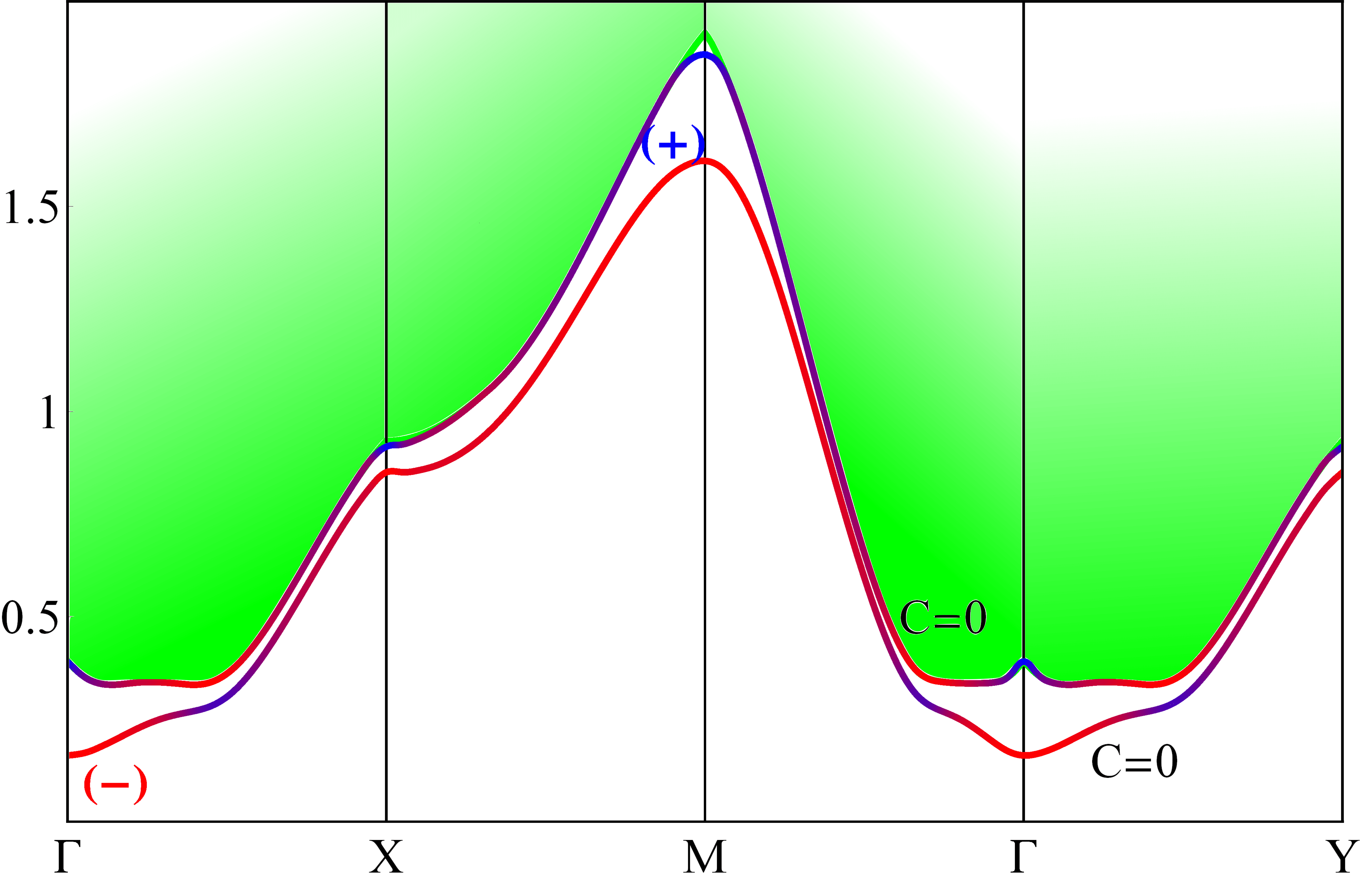}
\caption{$S_z=-1$ exciton dispersion along high-symmetry paths in the Brillouin zone, calculated for the two points E, top panel, and F, bottom panel, in Fig.~\ref{Fig5}. The green shaded regions are the particle-hole continuum. The blue and red colours of the curves indicate even (+) and odd (-) parity under inversion, while $C$ is the corresponding Chern number.}
\label{Fig6}
\end{figure}

In Fig.~\ref{Fig5} we show the Chern number of the lowest energy 
exciton branch with 
$S_z=-1$ calculated through Eq.~\eqn{eq-Chern} with $U_s=U_p=V=1.5$ 
as function of $M$ and $J$ along the way from the 
QSHI to the symmetry broken phase where excitons condense. We observe 
that the dipole-dipole interaction $J$ favours not only the instability of the $S_z=\pm 1$ excitons, but also their non trivial topology, signalled by a non zero Chern number. In Fig.~\ref{Fig6} we show for the 
two points E and F in Fig.~\ref{Fig5} the $S_z=-1$ exciton bands,  
$\omega_i(\bq)$, $i=1,2$ along high-symmetry paths in the BZ, together with the continuum of $S_z=-1$ particle-hole excitations, bounded from below by $\omega_\text{min}(\bq)$, see Eq.~\eqn{omega_i omega_min}. 
The upper branch is very lightly bound, and almost touch the continuum, 
unlike the lower branch, whose binding energy is maximum at the $\bd{\Gamma}$ point where, eventually, the condensation will take place. The blue and red colours indicate, respectively, even (+) and odd (-) parity character under inversion. We note that at point F in Fig.~\ref{Fig5} both 
exciton bands have vanishing Chern number, signalled by the same parity character at all inversion-invariant $\bk$-points. On the contrary, 
at point E, close to the transition, the two exciton branches change parity character among the high-symmetry points, and thus acquire finite 
and opposite Chern numbers, $C=\pm 2$. \\

\begin{figure}[hbt]
\centering
\includegraphics[width=0.75\columnwidth]{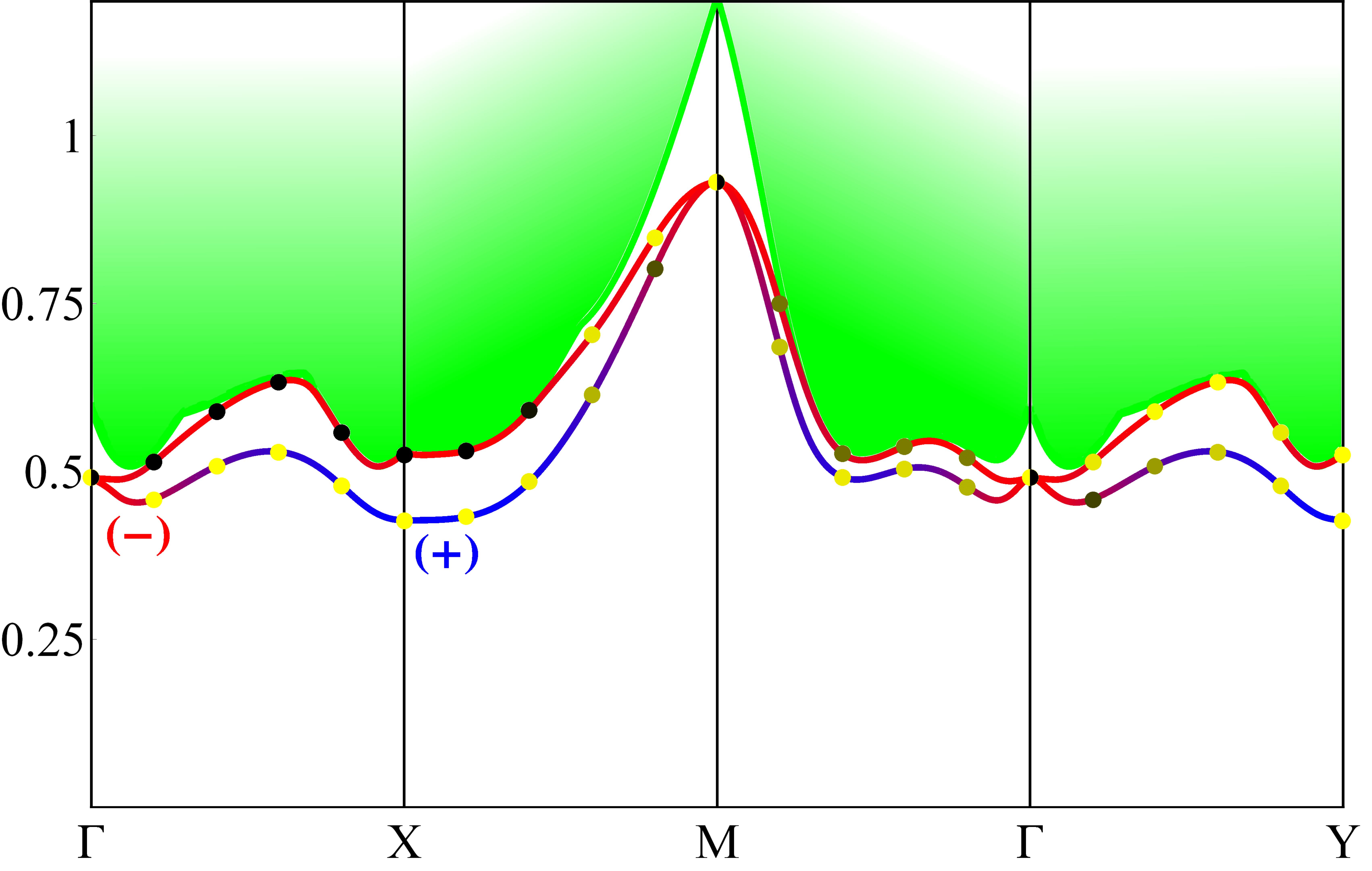}\\
\includegraphics[width=0.75\columnwidth]{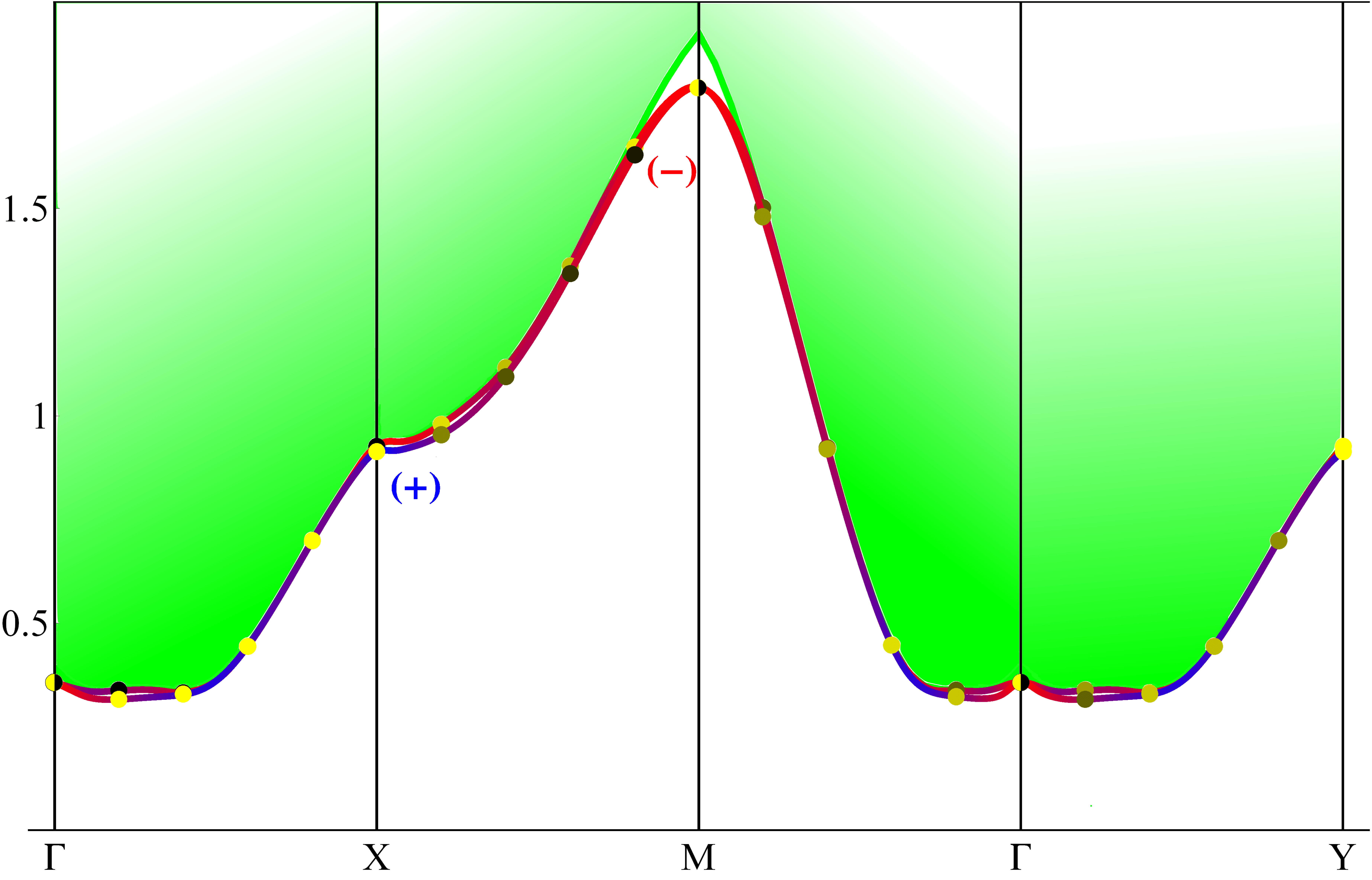}
\caption{Same as Fig.~\ref{Fig6} but for the $S_z=0$ excitons. Black and yellow dots indicate that the excitons are, respectively, even and odd under time reversal. }
\label{Fig7}
\end{figure}

For completeness, in Fig.~\ref{Fig7} we show at the same points E and F of Fig.~\ref{Fig5} the dispersion of the $S_z=0$ excitons. Since they are invariant 
under time reversal, we also indicate their symmetry, even (black dots) or odd (yellow dots), 
which correspond, respectively, to the spin singlet and spin triplet with $S_z=0$ components of the exciton. \\
Comparing Fig.~\ref{Fig7} with \ref{Fig6}, we note that the $S_z=0$ excitons are far  
less bound than the $S_z=-1$ ones. However, it is conceivable that the inclusion of the long range part of the Coulomb interaction could increase the binding energy of the $S_z=0$ excitons, even though we believe that the $S_z=\pm 1$ excitons will still be lower in energy.

Moving to the sample surface at point E we expect two phenomena to occur. First, chiral 
exciton edge modes should appear, and connect the two branches with opposite Chern numbers, in analogy with the single particle case, and as thoroughly discussed by the authors of  Ref.~\onlinecite{PhysRevB.96.161101} in the magnetised BHZ model. In addition, our previous results in the ribbon geometry, showing that the exciton condensate appears on the surface earlier than the bulk, suggest the existence of genuine surface excitons, more bound than their bulk counterparts, definitely in the $S_z=\pm 1$ channel, but possibly also in the $S_z=0$ one.\\
Both the chiral exciton edge modes as well as the surface excitons 
may potentially have important effects on the physical behaviour at the 
boundaries. First of all, since the most bound ones correspond to coherent $S_z=\pm 1$ 
particle-hole excitations, they may provide efficient decay channels for 
the single-particle edge modes, which are counter propagating waves 
with opposite $S_z=\pm 1/2$. Experimental evidences of such phenomenon 
in the purported topological Kondo insulator $\text{Sm}\text{B}_6$ 
have been indeed observed \cite{Park6599, PhysRevB.94.235125}, 
and previously attributed to scattering off bulk excitons \cite{PhysRevB.92.085133}. This is well possible, but should be much less
efficient than the scattering off surface exciton modes, which we propose as an alternative explanation. Furthermore, the presence of odd-parity excitons localised at the surface might have direct consequences on the surface optical activity, which could be worth investigating.

\subsection{Exciton condensate and magnetoelectricity}
\label{Results:C}

Since the order parameter in the phase with exciton condensation breaks 
spin $U(1)$ symmetry, inversion $\mathcal{I}$ and time reversal $\mathcal{T}$, but not 
$\mathcal{T}\times\mathcal{I}$, it is eligible to display 
magnetoelectric effects, which can be experimentally detected.\\ 
The free energy density expanded up to second order in the external electric and magnetic fields, both assumed constant in space and time, 
can be written as  
\beal
F(\mathbf{E},\mathbf{B}) = F_0 & - 
\fract{1}{2}\,\bd{E}\cdot\hat{\chi}_e\,\bd{E}
- \fract{1}{2}\,\bd{B}\cdot\hat{\chi}\,\bd{B}\\
&\qquad - \bd{E}\cdot\hat{\alpha}\,\bd{B}\,,
\eal
where $\hat{\chi}_e$, $\hat{\chi}$ and $\hat{\alpha}$ are 
the electric polarisability, magnetic susceptibility, and magnetoelectric tensors, respectively. 
The magnetization, $\bd{M}$, and polarization, $\bd{P}$,  are conjugate variables of the fields, namely 
\beal
\bd{M} &= - \frac{\partial F}{\partial \bd{B}}  =  
\hat{\chi}\,\bd{B} + \hat{\alpha}\,\bd{E}
\,,\\
\bd{P} &= - \frac{\partial F}{\partial \bd{E}}  =  
\hat{\chi}_e\,\bd{E} + \hat{\alpha}\,\bd{B}\, .
\eal
We observe that, since $\bd{E}$ and $\bd{B}$ have opposite properties under inversion and time reversal, a non-zero $\hat{\alpha}$ is allowed only when both symmetries are broken, 
but not their product.  
\\
Since the exciton condensate Eq.~\eqn{Delta}   
is spin-polarised in the $x\!-\!y$ plane, with azimuthal angle $\phi$, 
and involves dipole excitations $s\leftrightarrow p_{\pm 1}$, see
Eq.~\eqn{p-orbitals}, we restrict our analysis to fields $\bd{E}$ and $\bd{B}$ that have only $x$ and $y$ components, which allows us discarding 
the electromagnetic coupling to the electron charge current. 
Consequently, the magnetoelectric tensor $\hat{\alpha}$ of our interest will be a $2\times 2$ matrix with components $\alpha_{ij}$, $i,j=x,y$.

\begin{figure}[thb]
\includegraphics[width=0.8\columnwidth]{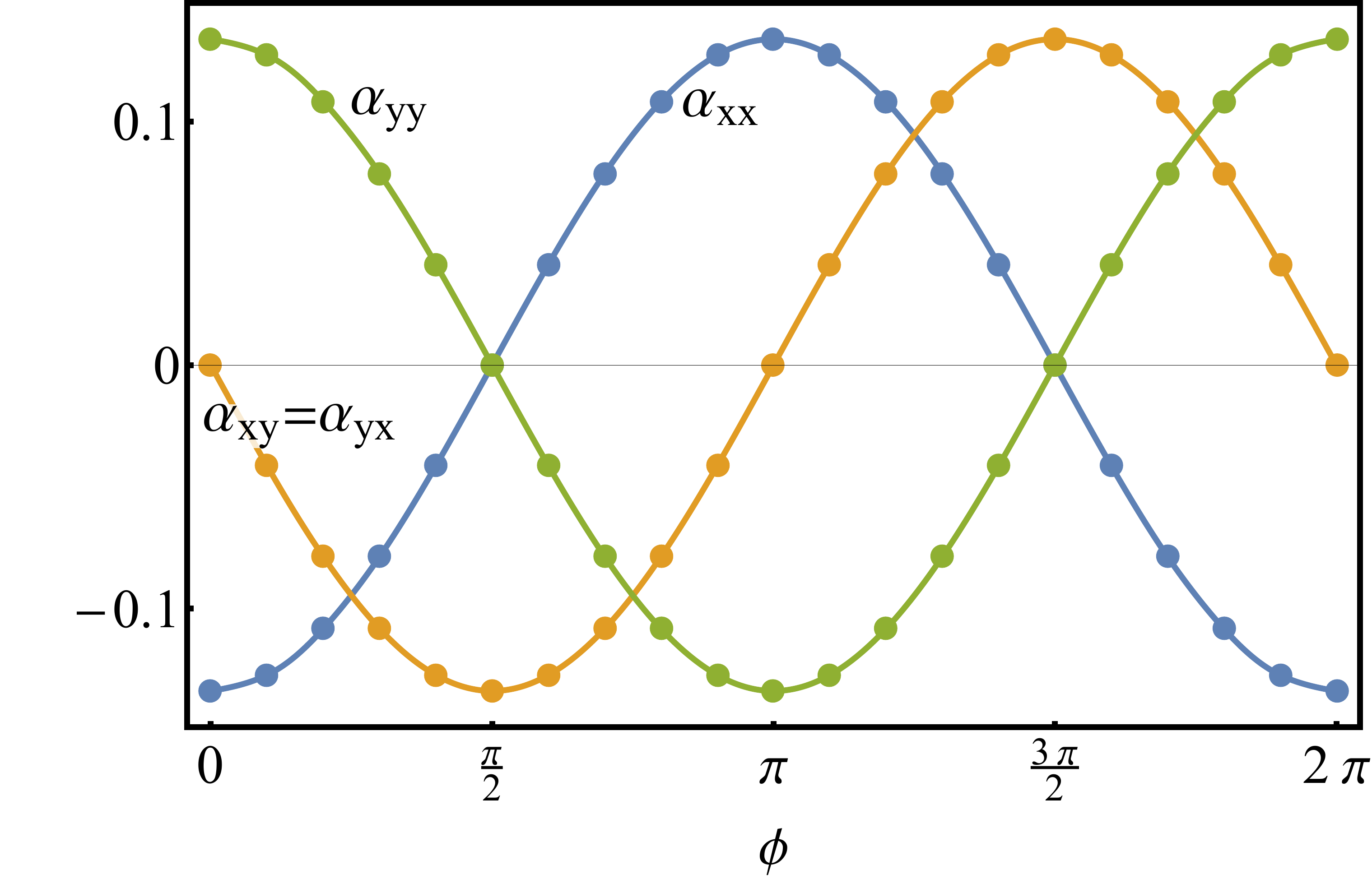}
\caption{Components of the magnetoelectric tensor $\hat{\alpha}$ calculated at the point D in Fig.~\ref{Fig2} ($M=0.5$, $U=2$). The data fit perfectly with the expression in Eq.~\eqn{alpha-fit}. }
\label{Fig8}
\end{figure}

In the exciton condensed phase, which is insulating, the coupling to the planar electric field is via the polarisation density, namely, in proper units,   
\beal
\delta H_E &= -\sum_i\, \bd{\Psi}_i^\dagger\,
\Big(\, E_x\,\hat{d}_x + E_y\,\hat{d}_y\,\Big)\,\bd{\Psi}^\dagga_i\,,
\eal
with dipole operators  
\beal
\hat{d}_x &= \sigma_0 \otimes \tau_1\,,& 
\hat{d}_y &= \sigma_3 \otimes \tau_2 \, . 
\eal
Moreover, as the orbitals $\ket{s\,\sigma}$ have physical total  
momentum $J_z = L_z+S_z=\pm 1/2$, while $\ket{p\,\sigma}$ have $J_z=\pm 3/2$, the in-plane magnetic field only couples to the magnetic moment of the $s$-orbitals. Specifically,  
\beal
\delta H_B &= -\sum_i\, \bd{\Psi}_i^\dagger\,
\Big(\, B_x\,\hat{m}_x + B_y\,\hat{m}_y\,\Big)\,\bd{\Psi}^\dagga_i\,,
\eal
where 
\beal
\hat{m}_x &= \sigma_1 \otimes \frac{\tau_0+\tau_3}{2}\;,&
\hat{m}_y &= \sigma_2 \otimes \frac{\tau_0+\tau_3}{2}\;.
\eal
Since we are interested in the effects of the external fields once the 
the symmetry has been broken, we performed a non self-consistent calculation with the HF self-energy calculated at $\bd{E}=\bd{B}
=\bd{0}$. The finite magnetoelectric effect in the presence of the exciton condensate is indeed confirmed, see Fig.~\ref{Fig8} where we show 
the components of $\hat{\alpha}$ as function of the azimuthal angle $\phi$ in Eq.~\eqn{Delta}, and which we find to behave as  
\be
\hat{\alpha}= \alpha_0 \begin{pmatrix}
-\cos \phi &  -\sin \phi \\ 
-\sin \phi &   \cos \phi
\end{pmatrix} \, ,\label{alpha-fit}
\ee
where $\alpha_0$ is proportional to the amplitude $\Delta$ 
of the order parameter, see Eq.~\eqn{Delta}, and thus vanishes when the 
symmetry is restored. \\
We remark that the magnetoelectric tensor \eqn{alpha-fit} has the form predicted 
for the magnetic point group $4'$~\cite{symmetry-alpha}, 
thus not showing signals of the nematic order proposed in Ref.~\onlinecite{PhysRevLett.120.186802}, as we earlier discussed in section \ref{Results:A}.

\section{Conclusions}
\label{Conclusions}
In this work we have studied within a conserving mean-field scheme the role of a local electron repulsion in the prototype BHZ model of a quantum spin Hall insulator 
\cite{Bernevig1757}, whose symmetries allows, besides the conventional monopole components of Coulomb interaction, also   
a dipolar one, which we find to play a rather important role. \\
In absence of interaction, the BHZ model displays, as function of a mass parameter $M>0$, two insulating phases, one topological at $M<M_c$, and another non-topological above $M_c$, 
separated by a metal point with Dirac-like dispersion at $M=M_c$. The primary effect of Coulomb interaction, namely the level repulsion between occupied and unoccupied states, pushes the critical $M_c$ to lower values, thus enlarging the stability region 
of the non-topological insulator. Besides that, and for intermediate values of $M$, our mean field results predict that interaction makes a new insulating phase to intrude between the topological and non-topological insulators, uncovering a path connecting the latter two that 
does not cross any metal point. In this phase, inversion symmetry and time reversal are spontaneously broken, though their product is not, implying the existence of magnetoelectric effects. The approach to this phase from both topological and non-topological sides is 
signalled by the softening of two exciton branches, related to each other by time-reversal 
and possessing, for $M\lesssim 1$ with the parameters of Fig.~\ref{Fig2}, finite and opposite Chern numbers.  
This phase can therefore be legitimately regarded as a condensate of topological excitons. \\
Since, starting from the quantum spin Hall insulator, the softening of those excitons and their eventual condensation occurs upon increasing the interaction, it is rather natural to 
expect those phenomena to be enhanced at the surface layers. Indeed, the mean field approach in a ribbon geometry predicts the surface instability to precede the bulk one. Even though a genuine exciton condensation at the surface layer might be prevented by quantum and thermal fluctuations, still it would sensibly affect the physics at the surface. The simplest consequence we may envisage is that the soft surface excitons would provide an efficient decay channel for the chiral single-particle edge modes, as indeed observed 
in the supposedly three-dimensional topological Kondo insulator $\text{Sm}\text{B}_6$~\cite{Park6599, PhysRevB.94.235125}. In addition, we cannot exclude important consequences on the transport properties and optical activity at the surface. 
\\
We believe that going beyond the approximations assumed throughout this paper should not significantly alter our main results. RPA plus exchange allows accessing in a simple way collective excitations, though it ignores their mutual interaction. We expect that the latter would surely affect the precise location of the transition points, but not wash out the exciton condensation. \\
Inclusion of the neglected long range tail of Coulomb interaction would introduce two terms: the standard monopole-monopole charge repulsion, proportional to $1/r$, and a dipole-dipole 
interaction decaying as $1/r^{3}$. The former is expected to increase the exciton binding energy, though without distinguishing between spin singlet and triplet channels. Therefore, our conclusion that the $S_z=\pm 1$ excitons soften earlier than the $S_z=0$ ones should remain even in presence of the $1/r$ tail of Coulomb interaction. The dipole-dipole interaction might instead favour an inhomogeneous exciton condensation. However, we suspect that the $1/r^3$ decay in two dimensions is not sufficient to stablize domains.  
\\
To conclude, we believe that our results, though obtained by a mean field calculation and for a specific model topological insulator, catch sight of still not fully explored effects of electron electron interaction in topological insulators, which might be worth investigating 
experimentally, as well as theoretically in other models and, eventually, by means of more reliable tools~\cite{Adriano-PRL2015,Adriano-PRB2016}.

\section*{Acknowledgments}
We are grateful to Francesca Paoletti, Massimo Capone, Fulvio Parmigiani and, especially, 
to Adriano Amaricci for useful discussions and comments. We also thank Fei Xue for drawing to our attention Ref.~\onlinecite{PhysRevLett.120.186802}. This work has been supported by the European Union under H2020 Framework Programs, ERC Advanced Grant No. 692670 ``FIRSTORM''.

\bibliographystyle{apsrev}
\bibliography{mybiblio}

\begin{thebibliography}{43}
\expandafter\ifx\csname natexlab\endcsname\relax\def\natexlab#1{#1}\fi
\expandafter\ifx\csname bibnamefont\endcsname\relax
  \def\bibnamefont#1{#1}\fi
\expandafter\ifx\csname bibfnamefont\endcsname\relax
  \def\bibfnamefont#1{#1}\fi
\expandafter\ifx\csname citenamefont\endcsname\relax
  \def\citenamefont#1{#1}\fi
\expandafter\ifx\csname url\endcsname\relax
  \def\url#1{\texttt{#1}}\fi
\expandafter\ifx\csname urlprefix\endcsname\relax\def\urlprefix{URL }\fi
\providecommand{\bibinfo}[2]{#2}
\providecommand{\eprint}[2][]{\url{#2}}

\bibitem[{\citenamefont{Seradjeh et~al.}(2009)\citenamefont{Seradjeh, Moore,
  and Franz}}]{Moore-exciton-PRL2009}
\bibinfo{author}{\bibfnamefont{B.}~\bibnamefont{Seradjeh}},
  \bibinfo{author}{\bibfnamefont{J.~E.} \bibnamefont{Moore}}, \bibnamefont{and}
  \bibinfo{author}{\bibfnamefont{M.}~\bibnamefont{Franz}},
  \bibinfo{journal}{Phys. Rev. Lett.} \textbf{\bibinfo{volume}{103}},
  \bibinfo{pages}{066402} (\bibinfo{year}{2009}),
  \urlprefix\url{https://link.aps.org/doi/10.1103/PhysRevLett.103.066402}.

\bibitem[{\citenamefont{Pikulin and Hyart}(2014)}]{Pikulin-PRL2014}
\bibinfo{author}{\bibfnamefont{D.~I.} \bibnamefont{Pikulin}} \bibnamefont{and}
  \bibinfo{author}{\bibfnamefont{T.}~\bibnamefont{Hyart}},
  \bibinfo{journal}{Phys. Rev. Lett.} \textbf{\bibinfo{volume}{112}},
  \bibinfo{pages}{176403} (\bibinfo{year}{2014}),
  \urlprefix\url{https://link.aps.org/doi/10.1103/PhysRevLett.112.176403}.

\bibitem[{\citenamefont{Budich et~al.}(2014)\citenamefont{Budich, Trauzettel,
  and Michetti}}]{Budich-PRL2014}
\bibinfo{author}{\bibfnamefont{J.~C.} \bibnamefont{Budich}},
  \bibinfo{author}{\bibfnamefont{B.}~\bibnamefont{Trauzettel}},
  \bibnamefont{and} \bibinfo{author}{\bibfnamefont{P.}~\bibnamefont{Michetti}},
  \bibinfo{journal}{Phys. Rev. Lett.} \textbf{\bibinfo{volume}{112}},
  \bibinfo{pages}{146405} (\bibinfo{year}{2014}),
  \urlprefix\url{https://link.aps.org/doi/10.1103/PhysRevLett.112.146405}.

\bibitem[{\citenamefont{Fuhrman et~al.}(2015)\citenamefont{Fuhrman, Leiner,
  Nikoli\ifmmode~\acute{c}\else \'{c}\fi{}, Granroth, Stone, Lumsden,
  DeBeer-Schmitt, Alekseev, Mignot, Koohpayeh
  et~al.}}]{Fuhrman-exciton-PRL2015}
\bibinfo{author}{\bibfnamefont{W.~T.} \bibnamefont{Fuhrman}},
  \bibinfo{author}{\bibfnamefont{J.}~\bibnamefont{Leiner}},
  \bibinfo{author}{\bibfnamefont{P.}~\bibnamefont{Nikoli\ifmmode~\acute{c}\else
  \'{c}\fi{}}}, \bibinfo{author}{\bibfnamefont{G.~E.} \bibnamefont{Granroth}},
  \bibinfo{author}{\bibfnamefont{M.~B.} \bibnamefont{Stone}},
  \bibinfo{author}{\bibfnamefont{M.~D.} \bibnamefont{Lumsden}},
  \bibinfo{author}{\bibfnamefont{L.}~\bibnamefont{DeBeer-Schmitt}},
  \bibinfo{author}{\bibfnamefont{P.~A.} \bibnamefont{Alekseev}},
  \bibinfo{author}{\bibfnamefont{J.-M.} \bibnamefont{Mignot}},
  \bibinfo{author}{\bibfnamefont{S.~M.} \bibnamefont{Koohpayeh}},
  \bibnamefont{et~al.}, \bibinfo{journal}{Phys. Rev. Lett.}
  \textbf{\bibinfo{volume}{114}}, \bibinfo{pages}{036401}
  (\bibinfo{year}{2015}),
  \urlprefix\url{https://link.aps.org/doi/10.1103/PhysRevLett.114.036401}.

\bibitem[{\citenamefont{Park et~al.}(2016)\citenamefont{Park, Sun, Noddings,
  Kim, Fisk, and Greene}}]{Park6599}
\bibinfo{author}{\bibfnamefont{W.~K.} \bibnamefont{Park}},
  \bibinfo{author}{\bibfnamefont{L.}~\bibnamefont{Sun}},
  \bibinfo{author}{\bibfnamefont{A.}~\bibnamefont{Noddings}},
  \bibinfo{author}{\bibfnamefont{D.-J.} \bibnamefont{Kim}},
  \bibinfo{author}{\bibfnamefont{Z.}~\bibnamefont{Fisk}}, \bibnamefont{and}
  \bibinfo{author}{\bibfnamefont{L.~H.} \bibnamefont{Greene}},
  \bibinfo{journal}{Proceedings of the National Academy of Sciences}
  \textbf{\bibinfo{volume}{113}}, \bibinfo{pages}{6599} (\bibinfo{year}{2016}),
  ISSN \bibinfo{issn}{0027-8424},
  \eprint{https://www.pnas.org/content/113/24/6599.full.pdf},
  \urlprefix\url{https://www.pnas.org/content/113/24/6599}.

\bibitem[{\citenamefont{Knolle and Cooper}(2017)}]{PhysRevLett.118.096604}
\bibinfo{author}{\bibfnamefont{J.}~\bibnamefont{Knolle}} \bibnamefont{and}
  \bibinfo{author}{\bibfnamefont{N.~R.} \bibnamefont{Cooper}},
  \bibinfo{journal}{Phys. Rev. Lett.} \textbf{\bibinfo{volume}{118}},
  \bibinfo{pages}{096604} (\bibinfo{year}{2017}),
  \urlprefix\url{https://link.aps.org/doi/10.1103/PhysRevLett.118.096604}.

\bibitem[{\citenamefont{Du et~al.}(2017)\citenamefont{Du, Li, Lou, Sullivan,
  Chang, Kono, and Du}}]{Du-Nat-Comm-2017}
\bibinfo{author}{\bibfnamefont{L.}~\bibnamefont{Du}},
  \bibinfo{author}{\bibfnamefont{X.}~\bibnamefont{Li}},
  \bibinfo{author}{\bibfnamefont{W.}~\bibnamefont{Lou}},
  \bibinfo{author}{\bibfnamefont{G.}~\bibnamefont{Sullivan}},
  \bibinfo{author}{\bibfnamefont{K.}~\bibnamefont{Chang}},
  \bibinfo{author}{\bibfnamefont{J.}~\bibnamefont{Kono}}, \bibnamefont{and}
  \bibinfo{author}{\bibfnamefont{R.-R.} \bibnamefont{Du}},
  \bibinfo{journal}{Nature Communications} \textbf{\bibinfo{volume}{8}},
  \bibinfo{pages}{1971} (\bibinfo{year}{2017}),
  \urlprefix\url{https://doi.org/10.1038/s41467-017-01988-1}.

\bibitem[{\citenamefont{Kung et~al.}(2019)\citenamefont{Kung, Goyal, Maslov,
  Wang, Lee, Kemper, Cheong, and Blumberg}}]{Bi2Se3}
\bibinfo{author}{\bibfnamefont{S.}~\bibnamefont{Kung}},
  \bibinfo{author}{\bibfnamefont{A.}~\bibnamefont{Goyal}},
  \bibinfo{author}{\bibfnamefont{D.}~\bibnamefont{Maslov}},
  \bibinfo{author}{\bibfnamefont{X.}~\bibnamefont{Wang}},
  \bibinfo{author}{\bibfnamefont{A.}~\bibnamefont{Lee}},
  \bibinfo{author}{\bibfnamefont{A.}~\bibnamefont{Kemper}},
  \bibinfo{author}{\bibfnamefont{S.-W.} \bibnamefont{Cheong}},
  \bibnamefont{and} \bibinfo{author}{\bibfnamefont{G.}~\bibnamefont{Blumberg}},
  \bibinfo{journal}{Proceedings of the National Academy of Sciences}
  \textbf{\bibinfo{volume}{116}}, \bibinfo{pages}{201813514}
  (\bibinfo{year}{2019}).

\bibitem[{\citenamefont{Varsano et~al.}(2020)\citenamefont{Varsano, Palummo,
  Molinari, and Rontani}}]{Molinari-Nature-Nano-2020}
\bibinfo{author}{\bibfnamefont{D.}~\bibnamefont{Varsano}},
  \bibinfo{author}{\bibfnamefont{M.}~\bibnamefont{Palummo}},
  \bibinfo{author}{\bibfnamefont{E.}~\bibnamefont{Molinari}}, \bibnamefont{and}
  \bibinfo{author}{\bibfnamefont{M.}~\bibnamefont{Rontani}},
  \bibinfo{journal}{Nature Nanotechnology} \textbf{\bibinfo{volume}{15}},
  \bibinfo{pages}{367} (\bibinfo{year}{2020}),
  \urlprefix\url{https://doi.org/10.1038/s41565-020-0650-4}.

\bibitem[{\citenamefont{Qian et~al.}(2014)\citenamefont{Qian, Liu, Fu, and
  Li}}]{Qian-TMDC-Science2014}
\bibinfo{author}{\bibfnamefont{X.}~\bibnamefont{Qian}},
  \bibinfo{author}{\bibfnamefont{J.}~\bibnamefont{Liu}},
  \bibinfo{author}{\bibfnamefont{L.}~\bibnamefont{Fu}}, \bibnamefont{and}
  \bibinfo{author}{\bibfnamefont{J.}~\bibnamefont{Li}},
  \bibinfo{journal}{Science} \textbf{\bibinfo{volume}{346}},
  \bibinfo{pages}{1344} (\bibinfo{year}{2014}), ISSN \bibinfo{issn}{0036-8075},
  \eprint{https://science.sciencemag.org/content/346/6215/1344.full.pdf},
  \urlprefix\url{https://science.sciencemag.org/content/346/6215/1344}.

\bibitem[{\citenamefont{Tang et~al.}(2017)\citenamefont{Tang, Zhang, Wong,
  Pedramrazi, Tsai, Jia, Moritz, Claassen, Ryu, Kahn
  et~al.}}]{Tang-Nat-Phys-2017}
\bibinfo{author}{\bibfnamefont{S.}~\bibnamefont{Tang}},
  \bibinfo{author}{\bibfnamefont{C.}~\bibnamefont{Zhang}},
  \bibinfo{author}{\bibfnamefont{D.}~\bibnamefont{Wong}},
  \bibinfo{author}{\bibfnamefont{Z.}~\bibnamefont{Pedramrazi}},
  \bibinfo{author}{\bibfnamefont{H.-Z.} \bibnamefont{Tsai}},
  \bibinfo{author}{\bibfnamefont{C.}~\bibnamefont{Jia}},
  \bibinfo{author}{\bibfnamefont{B.}~\bibnamefont{Moritz}},
  \bibinfo{author}{\bibfnamefont{M.}~\bibnamefont{Claassen}},
  \bibinfo{author}{\bibfnamefont{H.}~\bibnamefont{Ryu}},
  \bibinfo{author}{\bibfnamefont{S.}~\bibnamefont{Kahn}}, \bibnamefont{et~al.},
  \bibinfo{journal}{Nature Physics} \textbf{\bibinfo{volume}{13}},
  \bibinfo{pages}{683} (\bibinfo{year}{2017}),
  \urlprefix\url{https://doi.org/10.1038/nphys4174}.

\bibitem[{\citenamefont{Peng et~al.}(2017)\citenamefont{Peng, Yuan, Li, Yang,
  Xian, Yi, Shi, and Fu}}]{Peng-Nat-Comm-2017}
\bibinfo{author}{\bibfnamefont{L.}~\bibnamefont{Peng}},
  \bibinfo{author}{\bibfnamefont{Y.}~\bibnamefont{Yuan}},
  \bibinfo{author}{\bibfnamefont{G.}~\bibnamefont{Li}},
  \bibinfo{author}{\bibfnamefont{X.}~\bibnamefont{Yang}},
  \bibinfo{author}{\bibfnamefont{J.-J.} \bibnamefont{Xian}},
  \bibinfo{author}{\bibfnamefont{C.-J.} \bibnamefont{Yi}},
  \bibinfo{author}{\bibfnamefont{Y.-G.} \bibnamefont{Shi}}, \bibnamefont{and}
  \bibinfo{author}{\bibfnamefont{Y.-S.} \bibnamefont{Fu}},
  \bibinfo{journal}{Nature Communications} \textbf{\bibinfo{volume}{8}},
  \bibinfo{pages}{659} (\bibinfo{year}{2017}),
  \urlprefix\url{https://doi.org/10.1038/s41467-017-00745-8}.

\bibitem[{\citenamefont{Efimkin et~al.}(2012)\citenamefont{Efimkin, Lozovik,
  and Sokolik}}]{Efimkin-PRB2012}
\bibinfo{author}{\bibfnamefont{D.~K.} \bibnamefont{Efimkin}},
  \bibinfo{author}{\bibfnamefont{Y.~E.} \bibnamefont{Lozovik}},
  \bibnamefont{and} \bibinfo{author}{\bibfnamefont{A.~A.}
  \bibnamefont{Sokolik}}, \bibinfo{journal}{Phys. Rev. B}
  \textbf{\bibinfo{volume}{86}}, \bibinfo{pages}{115436}
  (\bibinfo{year}{2012}),
  \urlprefix\url{https://link.aps.org/doi/10.1103/PhysRevB.86.115436}.

\bibitem[{\citenamefont{Marchand and Franz}(2012)}]{Franz-thin-PRB2012}
\bibinfo{author}{\bibfnamefont{D.~J.~J.} \bibnamefont{Marchand}}
  \bibnamefont{and} \bibinfo{author}{\bibfnamefont{M.}~\bibnamefont{Franz}},
  \bibinfo{journal}{Phys. Rev. B} \textbf{\bibinfo{volume}{86}},
  \bibinfo{pages}{155146} (\bibinfo{year}{2012}),
  \urlprefix\url{https://link.aps.org/doi/10.1103/PhysRevB.86.155146}.

\bibitem[{\citenamefont{Mink et~al.}(2012)\citenamefont{Mink, Stoof, Duine,
  Polini, and Vignale}}]{Vignale-exciton-PRL2012}
\bibinfo{author}{\bibfnamefont{M.~P.} \bibnamefont{Mink}},
  \bibinfo{author}{\bibfnamefont{H.~T.~C.} \bibnamefont{Stoof}},
  \bibinfo{author}{\bibfnamefont{R.~A.} \bibnamefont{Duine}},
  \bibinfo{author}{\bibfnamefont{M.}~\bibnamefont{Polini}}, \bibnamefont{and}
  \bibinfo{author}{\bibfnamefont{G.}~\bibnamefont{Vignale}},
  \bibinfo{journal}{Phys. Rev. Lett.} \textbf{\bibinfo{volume}{108}},
  \bibinfo{pages}{186402} (\bibinfo{year}{2012}),
  \urlprefix\url{https://link.aps.org/doi/10.1103/PhysRevLett.108.186402}.

\bibitem[{\citenamefont{Lozovik and Sokolik}(2008)}]{Lozovik-BLG-JETP2008}
\bibinfo{author}{\bibfnamefont{Y.~E.} \bibnamefont{Lozovik}} \bibnamefont{and}
  \bibinfo{author}{\bibfnamefont{A.~A.} \bibnamefont{Sokolik}},
  \bibinfo{journal}{JETP Letters} \textbf{\bibinfo{volume}{87}},
  \bibinfo{pages}{55} (\bibinfo{year}{2008}),
  \urlprefix\url{https://doi.org/10.1134/S002136400801013X}.

\bibitem[{\citenamefont{Min et~al.}(2008)\citenamefont{Min, Bistritzer, Su, and
  MacDonald}}]{MacDonald-BLG-PRB2008}
\bibinfo{author}{\bibfnamefont{H.}~\bibnamefont{Min}},
  \bibinfo{author}{\bibfnamefont{R.}~\bibnamefont{Bistritzer}},
  \bibinfo{author}{\bibfnamefont{J.-J.} \bibnamefont{Su}}, \bibnamefont{and}
  \bibinfo{author}{\bibfnamefont{A.~H.} \bibnamefont{MacDonald}},
  \bibinfo{journal}{Phys. Rev. B} \textbf{\bibinfo{volume}{78}},
  \bibinfo{pages}{121401(R)} (\bibinfo{year}{2008}),
  \urlprefix\url{https://link.aps.org/doi/10.1103/PhysRevB.78.121401}.

\bibitem[{\citenamefont{Liu et~al.}(2017)\citenamefont{Liu, Watanabe,
  Taniguchi, Halperin, and Kim}}]{Liu-BLG-Nat-Phys2017}
\bibinfo{author}{\bibfnamefont{X.}~\bibnamefont{Liu}},
  \bibinfo{author}{\bibfnamefont{K.}~\bibnamefont{Watanabe}},
  \bibinfo{author}{\bibfnamefont{T.}~\bibnamefont{Taniguchi}},
  \bibinfo{author}{\bibfnamefont{B.~I.} \bibnamefont{Halperin}},
  \bibnamefont{and} \bibinfo{author}{\bibfnamefont{P.}~\bibnamefont{Kim}},
  \bibinfo{journal}{Nature Physics} \textbf{\bibinfo{volume}{13}},
  \bibinfo{pages}{746} (\bibinfo{year}{2017}),
  \urlprefix\url{https://doi.org/10.1038/nphys4116}.

\bibitem[{\citenamefont{Laurita et~al.}(2016)\citenamefont{Laurita, Morris,
  Koohpayeh, Rosa, Phelan, Fisk, McQueen, and Armitage}}]{PhysRevB.94.165154}
\bibinfo{author}{\bibfnamefont{N.~J.} \bibnamefont{Laurita}},
  \bibinfo{author}{\bibfnamefont{C.~M.} \bibnamefont{Morris}},
  \bibinfo{author}{\bibfnamefont{S.~M.} \bibnamefont{Koohpayeh}},
  \bibinfo{author}{\bibfnamefont{P.~F.~S.} \bibnamefont{Rosa}},
  \bibinfo{author}{\bibfnamefont{W.~A.} \bibnamefont{Phelan}},
  \bibinfo{author}{\bibfnamefont{Z.}~\bibnamefont{Fisk}},
  \bibinfo{author}{\bibfnamefont{T.~M.} \bibnamefont{McQueen}},
  \bibnamefont{and} \bibinfo{author}{\bibfnamefont{N.~P.}
  \bibnamefont{Armitage}}, \bibinfo{journal}{Phys. Rev. B}
  \textbf{\bibinfo{volume}{94}}, \bibinfo{pages}{165154}
  (\bibinfo{year}{2016}),
  \urlprefix\url{https://link.aps.org/doi/10.1103/PhysRevB.94.165154}.

\bibitem[{\citenamefont{Stankiewicz et~al.}(2019)\citenamefont{Stankiewicz,
  Evangelisti, Rosa, Schlottmann, and Fisk}}]{Fisk-SmB6-PRB2019}
\bibinfo{author}{\bibfnamefont{J.}~\bibnamefont{Stankiewicz}},
  \bibinfo{author}{\bibfnamefont{M.}~\bibnamefont{Evangelisti}},
  \bibinfo{author}{\bibfnamefont{P.~F.~S.} \bibnamefont{Rosa}},
  \bibinfo{author}{\bibfnamefont{P.}~\bibnamefont{Schlottmann}},
  \bibnamefont{and} \bibinfo{author}{\bibfnamefont{Z.}~\bibnamefont{Fisk}},
  \bibinfo{journal}{Phys. Rev. B} \textbf{\bibinfo{volume}{99}},
  \bibinfo{pages}{045138} (\bibinfo{year}{2019}),
  \urlprefix\url{https://link.aps.org/doi/10.1103/PhysRevB.99.045138}.

\bibitem[{\citenamefont{Hartstein et~al.}(2018)\citenamefont{Hartstein, Toews,
  Hsu, Zeng, Chen, Hatnean, Zhang, Nakamura, Padgett, Rodway-Gant
  et~al.}}]{Hartstein-SmB6-Nat-Phys2018}
\bibinfo{author}{\bibfnamefont{M.}~\bibnamefont{Hartstein}},
  \bibinfo{author}{\bibfnamefont{W.~H.} \bibnamefont{Toews}},
  \bibinfo{author}{\bibfnamefont{Y.~T.} \bibnamefont{Hsu}},
  \bibinfo{author}{\bibfnamefont{B.}~\bibnamefont{Zeng}},
  \bibinfo{author}{\bibfnamefont{X.}~\bibnamefont{Chen}},
  \bibinfo{author}{\bibfnamefont{M.~C.} \bibnamefont{Hatnean}},
  \bibinfo{author}{\bibfnamefont{Q.~R.} \bibnamefont{Zhang}},
  \bibinfo{author}{\bibfnamefont{S.}~\bibnamefont{Nakamura}},
  \bibinfo{author}{\bibfnamefont{A.~S.} \bibnamefont{Padgett}},
  \bibinfo{author}{\bibfnamefont{G.}~\bibnamefont{Rodway-Gant}},
  \bibnamefont{et~al.}, \bibinfo{journal}{Nature Physics}
  \textbf{\bibinfo{volume}{14}}, \bibinfo{pages}{166} (\bibinfo{year}{2018}),
  \urlprefix\url{https://doi.org/10.1038/nphys4295}.

\bibitem[{\citenamefont{Kapilevich et~al.}(2015)\citenamefont{Kapilevich,
  Riseborough, Gray, Gulacsi, Durakiewicz, and Smith}}]{PhysRevB.92.085133}
\bibinfo{author}{\bibfnamefont{G.~A.} \bibnamefont{Kapilevich}},
  \bibinfo{author}{\bibfnamefont{P.~S.} \bibnamefont{Riseborough}},
  \bibinfo{author}{\bibfnamefont{A.~X.} \bibnamefont{Gray}},
  \bibinfo{author}{\bibfnamefont{M.}~\bibnamefont{Gulacsi}},
  \bibinfo{author}{\bibfnamefont{T.}~\bibnamefont{Durakiewicz}},
  \bibnamefont{and} \bibinfo{author}{\bibfnamefont{J.~L.} \bibnamefont{Smith}},
  \bibinfo{journal}{Phys. Rev. B} \textbf{\bibinfo{volume}{92}},
  \bibinfo{pages}{085133} (\bibinfo{year}{2015}),
  \urlprefix\url{https://link.aps.org/doi/10.1103/PhysRevB.92.085133}.

\bibitem[{\citenamefont{Arab et~al.}(2016)\citenamefont{Arab, Gray,
  Nem\ifmmode~\check{s}\else \v{s}\fi{}\'ak, Evtushinsky, Schneider, Kim, Fisk,
  Rosa, Durakiewicz, and Riseborough}}]{PhysRevB.94.235125}
\bibinfo{author}{\bibfnamefont{A.}~\bibnamefont{Arab}},
  \bibinfo{author}{\bibfnamefont{A.~X.} \bibnamefont{Gray}},
  \bibinfo{author}{\bibfnamefont{S.}~\bibnamefont{Nem\ifmmode~\check{s}\else
  \v{s}\fi{}\'ak}}, \bibinfo{author}{\bibfnamefont{D.~V.}
  \bibnamefont{Evtushinsky}}, \bibinfo{author}{\bibfnamefont{C.~M.}
  \bibnamefont{Schneider}}, \bibinfo{author}{\bibfnamefont{D.-J.}
  \bibnamefont{Kim}}, \bibinfo{author}{\bibfnamefont{Z.}~\bibnamefont{Fisk}},
  \bibinfo{author}{\bibfnamefont{P.~F.~S.} \bibnamefont{Rosa}},
  \bibinfo{author}{\bibfnamefont{T.}~\bibnamefont{Durakiewicz}},
  \bibnamefont{and} \bibinfo{author}{\bibfnamefont{P.~S.}
  \bibnamefont{Riseborough}}, \bibinfo{journal}{Phys. Rev. B}
  \textbf{\bibinfo{volume}{94}}, \bibinfo{pages}{235125}
  (\bibinfo{year}{2016}),
  \urlprefix\url{https://link.aps.org/doi/10.1103/PhysRevB.94.235125}.

\bibitem[{\citenamefont{Akintola et~al.}(2018)\citenamefont{Akintola, Pal,
  R.~Dunsiger, C.~Y.~Fang, Potma, R.~Saha, Wang, Paglione, and
  E.~Sonier}}]{spinexciton}
\bibinfo{author}{\bibfnamefont{K.}~\bibnamefont{Akintola}},
  \bibinfo{author}{\bibfnamefont{A.}~\bibnamefont{Pal}},
  \bibinfo{author}{\bibfnamefont{S.}~\bibnamefont{R.~Dunsiger}},
  \bibinfo{author}{\bibfnamefont{A.}~\bibnamefont{C.~Y.~Fang}},
  \bibinfo{author}{\bibfnamefont{M.}~\bibnamefont{Potma}},
  \bibinfo{author}{\bibfnamefont{S.}~\bibnamefont{R.~Saha}},
  \bibinfo{author}{\bibfnamefont{X.}~\bibnamefont{Wang}},
  \bibinfo{author}{\bibfnamefont{J.}~\bibnamefont{Paglione}}, \bibnamefont{and}
  \bibinfo{author}{\bibfnamefont{J.}~\bibnamefont{E.~Sonier}},
  \bibinfo{journal}{npj Quantum Materials} \textbf{\bibinfo{volume}{3}}
  (\bibinfo{year}{2018}).

\bibitem[{\citenamefont{Garate and Franz}(2011)}]{Garate-exciton-PRB2011}
\bibinfo{author}{\bibfnamefont{I.}~\bibnamefont{Garate}} \bibnamefont{and}
  \bibinfo{author}{\bibfnamefont{M.}~\bibnamefont{Franz}},
  \bibinfo{journal}{Phys. Rev. B} \textbf{\bibinfo{volume}{84}},
  \bibinfo{pages}{045403} (\bibinfo{year}{2011}),
  \urlprefix\url{https://link.aps.org/doi/10.1103/PhysRevB.84.045403}.

\bibitem[{\citenamefont{Chen and Shindou}(2017)}]{PhysRevB.96.161101}
\bibinfo{author}{\bibfnamefont{K.}~\bibnamefont{Chen}} \bibnamefont{and}
  \bibinfo{author}{\bibfnamefont{R.}~\bibnamefont{Shindou}},
  \bibinfo{journal}{Phys. Rev. B} \textbf{\bibinfo{volume}{96}},
  \bibinfo{pages}{161101(R)} (\bibinfo{year}{2017}),
  \urlprefix\url{https://link.aps.org/doi/10.1103/PhysRevB.96.161101}.

\bibitem[{\citenamefont{Allocca et~al.}(2018)\citenamefont{Allocca, Efimkin,
  and Galitski}}]{Galitski-exciton-PRB2018}
\bibinfo{author}{\bibfnamefont{A.~A.} \bibnamefont{Allocca}},
  \bibinfo{author}{\bibfnamefont{D.~K.} \bibnamefont{Efimkin}},
  \bibnamefont{and} \bibinfo{author}{\bibfnamefont{V.~M.}
  \bibnamefont{Galitski}}, \bibinfo{journal}{Phys. Rev. B}
  \textbf{\bibinfo{volume}{98}}, \bibinfo{pages}{045430}
  (\bibinfo{year}{2018}),
  \urlprefix\url{https://link.aps.org/doi/10.1103/PhysRevB.98.045430}.

\bibitem[{\citenamefont{Bernevig et~al.}(2006)\citenamefont{Bernevig, Hughes,
  and Zhang}}]{Bernevig1757}
\bibinfo{author}{\bibfnamefont{B.~A.} \bibnamefont{Bernevig}},
  \bibinfo{author}{\bibfnamefont{T.~L.} \bibnamefont{Hughes}},
  \bibnamefont{and} \bibinfo{author}{\bibfnamefont{S.-C.} \bibnamefont{Zhang}},
  \bibinfo{journal}{Science} \textbf{\bibinfo{volume}{314}},
  \bibinfo{pages}{1757} (\bibinfo{year}{2006}), ISSN \bibinfo{issn}{0036-8075},
  \eprint{https://science.sciencemag.org/content/314/5806/1757.full.pdf},
  \urlprefix\url{https://science.sciencemag.org/content/314/5806/1757}.

\bibitem[{\citenamefont{Altmeyer et~al.}(2016)\citenamefont{Altmeyer,
  Guterding, Hirschfeld, Maier, Valent\'{\i}, and Scalapino}}]{RPA+E}
\bibinfo{author}{\bibfnamefont{M.}~\bibnamefont{Altmeyer}},
  \bibinfo{author}{\bibfnamefont{D.}~\bibnamefont{Guterding}},
  \bibinfo{author}{\bibfnamefont{P.~J.} \bibnamefont{Hirschfeld}},
  \bibinfo{author}{\bibfnamefont{T.~A.} \bibnamefont{Maier}},
  \bibinfo{author}{\bibfnamefont{R.}~\bibnamefont{Valent\'{\i}}},
  \bibnamefont{and} \bibinfo{author}{\bibfnamefont{D.~J.}
  \bibnamefont{Scalapino}}, \bibinfo{journal}{Phys. Rev. B}
  \textbf{\bibinfo{volume}{94}}, \bibinfo{pages}{214515}
  (\bibinfo{year}{2016}),
  \urlprefix\url{https://link.aps.org/doi/10.1103/PhysRevB.94.214515}.

\bibitem[{\citenamefont{Hughes et~al.}(2011)\citenamefont{Hughes, Prodan, and
  Bernevig}}]{PhysRevB.83.245132}
\bibinfo{author}{\bibfnamefont{T.~L.} \bibnamefont{Hughes}},
  \bibinfo{author}{\bibfnamefont{E.}~\bibnamefont{Prodan}}, \bibnamefont{and}
  \bibinfo{author}{\bibfnamefont{B.~A.} \bibnamefont{Bernevig}},
  \bibinfo{journal}{Phys. Rev. B} \textbf{\bibinfo{volume}{83}},
  \bibinfo{pages}{245132} (\bibinfo{year}{2011}),
  \urlprefix\url{https://link.aps.org/doi/10.1103/PhysRevB.83.245132}.

\bibitem[{\citenamefont{Ezawa et~al.}(2013)\citenamefont{Ezawa, Tanaka, and
  Nagaosa}}]{Ezawa-SciRep2013}
\bibinfo{author}{\bibfnamefont{M.}~\bibnamefont{Ezawa}},
  \bibinfo{author}{\bibfnamefont{Y.}~\bibnamefont{Tanaka}}, \bibnamefont{and}
  \bibinfo{author}{\bibfnamefont{N.}~\bibnamefont{Nagaosa}},
  \bibinfo{journal}{Scientific Reports} \textbf{\bibinfo{volume}{3}},
  \bibinfo{pages}{2790} (\bibinfo{year}{2013}),
  \urlprefix\url{https://doi.org/10.1038/srep02790}.

\bibitem[{\citenamefont{Xue and MacDonald}(2018)}]{PhysRevLett.120.186802}
\bibinfo{author}{\bibfnamefont{F.}~\bibnamefont{Xue}} \bibnamefont{and}
  \bibinfo{author}{\bibfnamefont{A.~H.} \bibnamefont{MacDonald}},
  \bibinfo{journal}{Phys. Rev. Lett.} \textbf{\bibinfo{volume}{120}},
  \bibinfo{pages}{186802} (\bibinfo{year}{2018}),
  \urlprefix\url{https://link.aps.org/doi/10.1103/PhysRevLett.120.186802}.

\bibitem[{\citenamefont{Shitade et~al.}(2009)\citenamefont{Shitade, Katsura,
  Kune\ifmmode~\check{s}\else \v{s}\fi{}, Qi, Zhang, and
  Nagaosa}}]{Shitade-PRL2009}
\bibinfo{author}{\bibfnamefont{A.}~\bibnamefont{Shitade}},
  \bibinfo{author}{\bibfnamefont{H.}~\bibnamefont{Katsura}},
  \bibinfo{author}{\bibfnamefont{J.}~\bibnamefont{Kune\ifmmode~\check{s}\else
  \v{s}\fi{}}}, \bibinfo{author}{\bibfnamefont{X.-L.} \bibnamefont{Qi}},
  \bibinfo{author}{\bibfnamefont{S.-C.} \bibnamefont{Zhang}}, \bibnamefont{and}
  \bibinfo{author}{\bibfnamefont{N.}~\bibnamefont{Nagaosa}},
  \bibinfo{journal}{Phys. Rev. Lett.} \textbf{\bibinfo{volume}{102}},
  \bibinfo{pages}{256403} (\bibinfo{year}{2009}),
  \urlprefix\url{https://link.aps.org/doi/10.1103/PhysRevLett.102.256403}.

\bibitem[{\citenamefont{Medhi et~al.}(2012)\citenamefont{Medhi, Shenoy, and
  Krishnamurthy}}]{PhysRevB.85.235449}
\bibinfo{author}{\bibfnamefont{A.}~\bibnamefont{Medhi}},
  \bibinfo{author}{\bibfnamefont{V.~B.} \bibnamefont{Shenoy}},
  \bibnamefont{and} \bibinfo{author}{\bibfnamefont{H.~R.}
  \bibnamefont{Krishnamurthy}}, \bibinfo{journal}{Phys. Rev. B}
  \textbf{\bibinfo{volume}{85}}, \bibinfo{pages}{235449}
  (\bibinfo{year}{2012}),
  \urlprefix\url{https://link.aps.org/doi/10.1103/PhysRevB.85.235449}.

\bibitem[{\citenamefont{Amaricci et~al.}(2017)\citenamefont{Amaricci,
  Privitera, Petocchi, Capone, Sangiovanni, and
  Trauzettel}}]{Adriano-edge-PRB2017}
\bibinfo{author}{\bibfnamefont{A.}~\bibnamefont{Amaricci}},
  \bibinfo{author}{\bibfnamefont{L.}~\bibnamefont{Privitera}},
  \bibinfo{author}{\bibfnamefont{F.}~\bibnamefont{Petocchi}},
  \bibinfo{author}{\bibfnamefont{M.}~\bibnamefont{Capone}},
  \bibinfo{author}{\bibfnamefont{G.}~\bibnamefont{Sangiovanni}},
  \bibnamefont{and}
  \bibinfo{author}{\bibfnamefont{B.}~\bibnamefont{Trauzettel}},
  \bibinfo{journal}{Phys. Rev. B} \textbf{\bibinfo{volume}{95}},
  \bibinfo{pages}{205120} (\bibinfo{year}{2017}),
  \urlprefix\url{https://link.aps.org/doi/10.1103/PhysRevB.95.205120}.

\bibitem[{\citenamefont{Rothe et~al.}(2010)\citenamefont{Rothe, Reinthaler,
  Liu, Molenkamp, Zhang, and Hankiewicz}}]{Rothe_2010}
\bibinfo{author}{\bibfnamefont{D.~G.} \bibnamefont{Rothe}},
  \bibinfo{author}{\bibfnamefont{R.~W.} \bibnamefont{Reinthaler}},
  \bibinfo{author}{\bibfnamefont{C.-X.} \bibnamefont{Liu}},
  \bibinfo{author}{\bibfnamefont{L.~W.} \bibnamefont{Molenkamp}},
  \bibinfo{author}{\bibfnamefont{S.-C.} \bibnamefont{Zhang}}, \bibnamefont{and}
  \bibinfo{author}{\bibfnamefont{E.~M.} \bibnamefont{Hankiewicz}},
  \bibinfo{journal}{New Journal of Physics} \textbf{\bibinfo{volume}{12}},
  \bibinfo{pages}{065012} (\bibinfo{year}{2010}),
  \urlprefix\url{https://iopscience.iop.org/article/10.1088/1367-2630/12/6/065012/meta}.

\bibitem[{\citenamefont{Kane and Mele}(2005)}]{PhysRevLett.95.226801}
\bibinfo{author}{\bibfnamefont{C.~L.} \bibnamefont{Kane}} \bibnamefont{and}
  \bibinfo{author}{\bibfnamefont{E.~J.} \bibnamefont{Mele}},
  \bibinfo{journal}{Phys. Rev. Lett.} \textbf{\bibinfo{volume}{95}},
  \bibinfo{pages}{226801} (\bibinfo{year}{2005}),
  \urlprefix\url{https://link.aps.org/doi/10.1103/PhysRevLett.95.226801}.

\bibitem[{\citenamefont{Mong and Shivamoggi}(2011)}]{PhysRevB.83.125109}
\bibinfo{author}{\bibfnamefont{R.~S.~K.} \bibnamefont{Mong}} \bibnamefont{and}
  \bibinfo{author}{\bibfnamefont{V.}~\bibnamefont{Shivamoggi}},
  \bibinfo{journal}{Phys. Rev. B} \textbf{\bibinfo{volume}{83}},
  \bibinfo{pages}{125109} (\bibinfo{year}{2011}),
  \urlprefix\url{https://link.aps.org/doi/10.1103/PhysRevB.83.125109}.

\bibitem[{\citenamefont{Amaricci et~al.}(2018)\citenamefont{Amaricci, Valli,
  Sangiovanni, Trauzettel, and Capone}}]{PhysRevB.98.045133}
\bibinfo{author}{\bibfnamefont{A.}~\bibnamefont{Amaricci}},
  \bibinfo{author}{\bibfnamefont{A.}~\bibnamefont{Valli}},
  \bibinfo{author}{\bibfnamefont{G.}~\bibnamefont{Sangiovanni}},
  \bibinfo{author}{\bibfnamefont{B.}~\bibnamefont{Trauzettel}},
  \bibnamefont{and} \bibinfo{author}{\bibfnamefont{M.}~\bibnamefont{Capone}},
  \bibinfo{journal}{Phys. Rev. B} \textbf{\bibinfo{volume}{98}},
  \bibinfo{pages}{045133} (\bibinfo{year}{2018}),
  \urlprefix\url{https://link.aps.org/doi/10.1103/PhysRevB.98.045133}.

\bibitem[{\citenamefont{Borghi et~al.}(2009)\citenamefont{Borghi, Fabrizio, and
  Tosatti}}]{PhysRevLett.102.066806}
\bibinfo{author}{\bibfnamefont{G.}~\bibnamefont{Borghi}},
  \bibinfo{author}{\bibfnamefont{M.}~\bibnamefont{Fabrizio}}, \bibnamefont{and}
  \bibinfo{author}{\bibfnamefont{E.}~\bibnamefont{Tosatti}},
  \bibinfo{journal}{Phys. Rev. Lett.} \textbf{\bibinfo{volume}{102}},
  \bibinfo{pages}{066806} (\bibinfo{year}{2009}),
  \urlprefix\url{https://link.aps.org/doi/10.1103/PhysRevLett.102.066806}.

\bibitem[{\citenamefont{Rivera}(2009)}]{symmetry-alpha}
\bibinfo{author}{\bibfnamefont{J.~P.} \bibnamefont{Rivera}},
  \bibinfo{journal}{The European Physical Journal B}
  \textbf{\bibinfo{volume}{71}}, \bibinfo{pages}{299} (\bibinfo{year}{2009}),
  \urlprefix\url{https://doi.org/10.1140/epjb/e2009-00336-7}.

\bibitem[{\citenamefont{Amaricci et~al.}(2015)\citenamefont{Amaricci, Budich,
  Capone, Trauzettel, and Sangiovanni}}]{Adriano-PRL2015}
\bibinfo{author}{\bibfnamefont{A.}~\bibnamefont{Amaricci}},
  \bibinfo{author}{\bibfnamefont{J.~C.} \bibnamefont{Budich}},
  \bibinfo{author}{\bibfnamefont{M.}~\bibnamefont{Capone}},
  \bibinfo{author}{\bibfnamefont{B.}~\bibnamefont{Trauzettel}},
  \bibnamefont{and}
  \bibinfo{author}{\bibfnamefont{G.}~\bibnamefont{Sangiovanni}},
  \bibinfo{journal}{Phys. Rev. Lett.} \textbf{\bibinfo{volume}{114}},
  \bibinfo{pages}{185701} (\bibinfo{year}{2015}),
  \urlprefix\url{https://link.aps.org/doi/10.1103/PhysRevLett.114.185701}.

\bibitem[{\citenamefont{Amaricci et~al.}(2016)\citenamefont{Amaricci, Budich,
  Capone, Trauzettel, and Sangiovanni}}]{Adriano-PRB2016}
\bibinfo{author}{\bibfnamefont{A.}~\bibnamefont{Amaricci}},
  \bibinfo{author}{\bibfnamefont{J.~C.} \bibnamefont{Budich}},
  \bibinfo{author}{\bibfnamefont{M.}~\bibnamefont{Capone}},
  \bibinfo{author}{\bibfnamefont{B.}~\bibnamefont{Trauzettel}},
  \bibnamefont{and}
  \bibinfo{author}{\bibfnamefont{G.}~\bibnamefont{Sangiovanni}},
  \bibinfo{journal}{Phys. Rev. B} \textbf{\bibinfo{volume}{93}},
  \bibinfo{pages}{235112} (\bibinfo{year}{2016}),
  \urlprefix\url{https://link.aps.org/doi/10.1103/PhysRevB.93.235112}.

\end{thebibliography}

\end{document}